\documentclass[times,3p]{elsarticle}
\usepackage{amssymb}
\usepackage[applemac]{inputenc} 

\journal{Studies in History and Philosophy of Modern Physics}

\begin{document}

\begin{frontmatter}

\title{What have we learned from observational cosmology ?}
\author[apc]{J.-Ch. Hamilton}
\ead{hamilton@apc.univ-paris7.fr}
\address[apc]{APC, Astroparticule et Cosmologie, Universit\'{e} Paris Diderot, CNRS/IN2P3, CEA/Irfu, Observatoire de Paris, Sorbonne Paris Cit\'{e}, 10, rue Alice Domon \& L\'{e}onie Duquet, 75205 Paris Cedex 13, France}

\begin{abstract}
We review the observational foundations of the $\Lambda$CDM model, considered by most cosmologists as the standard model of cosmology. The Cosmological Principle, a key assumption of the model is shown to be verified with increasing accuracy. The fact that the Universe seems to have expanded from and hot and dense past is supported by many independent probes (galaxy redshifts, Cosmic Microwave Background, Big-Bang Nucleosynthesis and reionization). The explosion of detailed observations in the last few decades has allowed for precise measurements of the cosmological parameters within Friedman-Lemaître-Robertson-Walker cosmologies leading to the $\Lambda$CDM model: an apparently flat Universe, dominated by a cosmological constant, whose matter component is dominantly dark.  We describe and discuss the various observational probes that led to this conclusion and conclude that the $\Lambda$CDM model, although leaving a number of open questions concerning the deep nature of the constituents of the Universe, provides the best theoretical framework to explain the observations.
\end{abstract}

\begin{keyword}
Observational Comsology
\end{keyword}

\end{frontmatter}

\section{The $\Lambda$CDM model}
\subsection{Construction of the model}
The so-called $\Lambda$CDM model, also known as the concordance model, is a particular case -- with a rather well defined set of cosmological parameters -- of the larger class of Friedman-Lemaître-Robertson-Walker models based on the Big-Bang paradigm. These models correspond to solutions of the equations of General Relativity for isotropic and homogeneous universes. There are therefore two major assumptions at the basis of $\Lambda$CDM: i) gravitation is the force that governs the overall behavior of the Universe and is described by General Relativity ; ii) the Cosmological Principle, stating that the Universe is homogeneous and isotropic -- the same everywhere in average.

The first assumption -- the validity of General Relativity -- is well established in the sense that no obvious excursion from this theory has ever been observed. It is strongly supported by the many successes of General Relativity in predicting observations such as the precise motion of bodies in the solar system or gravitational lensing. In General Relativity, gravitational forces are interpreted as resulting from a curvature of the space--time metric, the curvature being produced by the presence of masses and energy. The masses free-fall along geodesics of this curved space--time, resulting in apparently accelerated motions.
Basically, the curvature of space--time dictates the motion of masses, while masses affect the curvature. This mass/curvature interplay is summarized in the famous Einstein's equations, a complex set of non--linear differential equations whose unknown is the metric of space--time. In the most general case, these equations cannot be easily solved, only very specific cases lead to analytical solutions. The most interesting case for cosmology is the case of a homogeneous Universe where the curvature, smoothed over large scales, is the same everywhere at a given time. This {\em Cosmological Principle} was at first an assumption (motivated by extending the Copernican principle to the whole Universe) as it was not originally strongly supported by much observations. It was the simplest in order to solve Einstein's equations and proved to be extremely fruitful. It has been later supported by observations when technology allowed one to investigate the large scale structure of the Universe (see section~\ref{cosmological_principle}).

With the Cosmological Principle, it is straightforward to obtain a general expression for the (large-scales smoothed) metric of the Universe known as the Friedman-Lemaître-Robertson-Walker (FLRW) metric. It describes a Universe where the two unknowns are a global scale factor (amplitude and time evolution) and the constant curvature of space--time. The dynamics of the scale factor is obtained from Einstein's equations which simplify to Friedman's equations. These are simple differential equations for the scale factor that depend on the content of the Universe parametrized by various cosmological parameters. The first one is $\Omega_m$, the average matter density (that can be subdivided into its relativistic (light) species and the non-relativistic (heavy) ones) that undergoes dilution with the scale factor as expected. This matter content includes the known matter particles from the Standard Model of Particle Physics and the mysterious Dark Matter (see section~\ref{dark_matter}). An other important component is $\Omega_\Lambda=\frac{\Lambda}{3H^2}$, where $H=\frac{\dot{a}}{a}$ is the Hubble expansion rate and $\Lambda$ is a constant that does not dilute with the scale factor. $\Lambda$ is known as the cosmological constant in its simplest version, but that can have a more complex description under the name of Dark Energy (see section~\ref{dark_energy}). Finally the average curvature of the Universe $\Omega_k$ is related to the previous ones through $\Omega_k=1-\Omega_m-\Omega_\Lambda$. It can be negative, corresponding to an infinite hyperbolic Universe, exactly null, corresponding to an infinite flat Euclidean Universe or positive, corresponding to a closed spherical Universe\footnote{The infinity or finiteness of these universes given here corresponds to the case of trivial simply connected topologies, in the more general case of multi-connected topologies, one can have closed Universe with any of these curvatures.}. The striking feature of the FLRW models is that they are dynamical, the scale factor increases or decreases with time but can never be kept constant in a stable manner. This therefore corresponds to expanding or contracting universes.

In the modern version of FLRW models, known as $\Lambda$CDM models, there are about 12 free parameters that includes the ones cited above plus the current rate of expansion and parameters describing the primordial density perturbations that gave rise to the large scale structure observed around us. These parameters are constrained by observations to a few percents (and even below) for most of them. The current expansion rate is around $72~\mathrm{km.s^{-1}.Mpc^{-1}}$ and is measured by a number of independent probes. Matter counts for around 26\% of the Universe, and only 4 of these 26\% are due to ordinary matter. The largest part, 22\% is due to matter that acts only through gravitation but does not interact with light, hence the name of Dark Matter. The 74 remaining percents of the energy content of the Universe are apparently due to Dark Energy. The total energy density (Matter + Dark Matter + Dark Energy) is compatible to within one percent with the critical one corresponding to a zero curvature Universe. Here again these numbers rely on complementary and concordant observations of the Cosmic Microwave Background (see section~\ref{cmb}), structure formation, distant type Ia supernovae and other probes.
With such a domination of Dark Energy, the fate of the Universe, as predicted by the Friedman equations, is eternal accelerated expansion\footnote{This is the case unless dark energy vanishes at some point, in such a case one is back to the $\Lambda=0$ situation where the fate of the Universe is determined by the value of the curvature.}.

If one trusts the initial assumptions of $\Lambda$CDM cosmology, there is hardly any space out of this scheme. One has to relax some of the assumptions to follow different paths such as refuting General Relativity as the correct theory of gravitation on cosmological scales (which cannot be currently excluded by observations), refuting the Cosmological Principle or its application to solving the equations for the scale factor of the Universe. There are strong observational constraints limiting these possibilities but they cannot be fully excluded as of today and these possibilities should therefore be kept in mind when interpreting the results of observational cosmology, especially as, although the observational success of $\Lambda$CDM model is extremely impressive, these observations lead to this surprising mixture in which 100\% are extremely well fitted by the model but only 4\% are really understood.

\subsection{Tests of the Cosmological Principle\label{cosmological_principle}}
As stated above, $\Lambda$CDM relies on the applicability of General Relativity on cosmological scales and on the use of the Cosmological Principle in order to make the calculations doable. While philosophically motivated by the Copernican principle, the latter argument, certainly important from a practical point of view, needs to be confronted with observations in order to be convincing.

Actually the term "Cosmological Principle" needs to be explained a little bit before describing the supporting observations. The idea of a homogeneous Universe is in fact an extension to the famous Copernican principle according to which the Earth is not at a privileged place in the Universe. In the cosmological context, the idea must be understood as "we are located in a region of the Universe that has nothing special, we could be living somewhere else and our observations, although not exactly the same at the anecdotic level, would be similar in average". It also comes with the supposition/observation that the Universe is isotropic: whatever the direction we look at on the sky, we observe similar things. Of course we are inside a galaxy that is shaped like a fried egg, so we observe more stars in the direction of the Galactic plane and even more in the direction of the Galactic center, but this is a local effect. If we look beyond, the isotropy was already quite convincing at the beginning of the 20th century, when this was established as a principle.

By adding the Copernican principle with the isotropy of the Universe, the global homogeneity follows. This can be easily understood with hand-waving arguments but is actually proven rigorously in the context of General Relativity \citep{maartens}. Imagine you are observing the sky from a point called P1 (your planet) and observe isotropy around you, you would then conclude that all points located at a given distance $r$ from you have the same properties (same density for instance). In particular point B and C located on this circle of radius $r$, one looking north and one looking south, have the same properties. Now imagine an alien observer living far from us in point P2 (his planet). If this alien astronomer also observes isotropy, when he looks at the same point B as you do, located at a distance $r'$ from him, should also see the same properties at point $C$ situated somewhere else on his circle of radius $r'$. As $A$ and $B$ have the same properties from your observations, and $B$ and $C$ also have the same properties from his observations, $A$ and $C$ (that are not at the same distance from neither of you nor the alien astronomer) must have the same properties. As this reasoning needs to be true for every circles and every locations in the Universe (isotropy and Copernican principle), the properties need to be the same at every location of the Universe (at a given time). The argument is of course true only for the largest scales where gravitational collapse did not induce bound structures. This is why isotropy and Copernican principle imply the Cosmological Principle "the Universe is homogeneous on the large scales".

\subsubsection{Isotropy}
Isotropy is nowadays well established throughout the observable Universe: using modern spectroscopic surveys, one can map the location of hundreds of thousands galaxies in three dimensions. We observe that they do cluster rather strongly due to gravitational attraction, resulting in sponge-like shapes with large voids separated by filaments of matter. But there is no evidence for a larger structure beyond scales of a few tens Mpc while we observe volumes up to about 2~Gpc$^3$, much larger than these clustering scales. The data collected by the SDSS-III/BOSS collaboration is the most up-to-date in this regard \citep{BOSS} and projected distributions on the celestial sphere of the locations of galaxies show an impressive isotropy in the galaxy distribution at redshifts between 0.5 and 0.6. Beyond these scales, the observation of the perfect black-body nature of the Cosmic Microwave Background radiation (see section~\ref{cmb}) shows a uniform temperature over the whole celestial sphere. The tiny temperature fluctuations that are observed (and well understood) are only of around 1/100000 of the average temperature, corresponding to density perturbations of the same order of magnitude.

\subsubsection{Copernican principle}
There is therefore a good set of observations pointing towards isotropy of the Universe. Now, although the Copernican principle seems "logical" or "well motivated" from a modest point of view, one must admit that the observational basis for it has not been as strong as often recognized until recently. The three-dimensional maps of the galaxy locations mentioned above provide good evidence that the Universe is close to be homogeneous out to redshifts of around 0.6, which is already a strong constraint, but one needs to explore homogeneity over further distances to fulfill the FLRW requirements at the basis of the $\Lambda$CDM model.

Criticisms on the assumption of the Copernican principle have been particularly active in the recent years as a possible explanation for the acceleration of the expansion -- alternative to the surprising dark energy (see section~\ref{dark_energy}) -- is that we live in a particularly under-dense region of the Universe (a large void), the faraway galaxies (beyond $z\sim 0.5$) falling towards the walls of this void. More generally, one could explain Dark Energy through the fact that we could be in a inhomogeneous Universe where the FLRW equations would be out of scope and could be replaced by the Lemaître-Tolman-Bondi \citep{ltb} space--time which could fit the accelerated expansion without the need of Dark Energy \citep{sarkar2010}. Actually it has been shown that in such models, centered on ourselves, the CMB seen by distant observers would be strongly inhomogeneous and should result in violations of the black-body nature of the CMB we observe \citep{stebbins}. It could indeed be isotropic, but inhomogeneity would induce spectral distortions that are not compatible with the observed spectrum of the CMB \citep{caldwell_stebbins} for the cases of large voids with density contrasts large enough to explain the acceleration of the expansion. Similarly, the CMB photons scattering on the hot ionized gas in distant galaxy clusters would exhibit a spectral distortion known as {\em Kinetic Sunyaev Zel'dovitch effect} (kSZ) \citep{sz1,sz2} that would be of appreciable magnitude if these clusters were in motion with respect to the CMB we observe, as one would expect in inhomogeneous universes. The lack of significant kSZ effect observed on a number of distant clusters \citep{zhang_stebbins} seems to disfavor the possibility of explaining the acceleration of the Universe by large scale inhomogeneities and therefore to validate the Copernican Principle. Future tests based on the consistency between distance and expansion rate measurements \citep{maartens} are expected to be available in the next few years using data from large volume galaxy surveys. In the long term, measurements of the time drift of the cosmological redshift \citep{uzan} should be the most powerful test to probe the homogeneity of the Universe, but these are not expected before many years as they require unprecedented redshift measurement accuracy.

Isotropy of the observed Universe and the Copernican Principle therefore both seem well tested enough that deviations from a homogeneous Universe and therefore from the FLRW cosmology should be smaller than the observed surprising acceleration of the expansion. Finally, it appears that there are no strong reasons to reject the applicability of the Cosmological Principle today \citep{jones} as it would not bring any simplification in the interpretation of the observations but would require a surprising specificity of our location in the Universe. There remains discussions on the possibility of a fractal distribution of the matter in the Universe \citep{gabrielli, pietronero} that seem to be hardly in agreement with recent studies of large volume galaxy catalogs where the transition towards homogeneity is observed beyond scales of $\sim 70 h^{-1}.\mathrm{Mpc}$ \citep{hogg, scrimgeour}.

\subsection{Tests of the $\Lambda$CDM paradigm}
The $\Lambda$CDM model is the updated version of the original Big-Bang model that include several amendments such as the presence of Cold Dark-Matter (the CDM), dark-energy or cosmological constant (the $\Lambda$) both motivated by observations as will be discussed below. The $\Lambda$CDM model also often implicitly includes the hypothesis of an inflationary phase in the early Universe, thought to be responsible for the generation of primordial perturbations, absent from the original Big-Bang model. The modern label for this model does not explicitly mention the "Big-Bang" as there is a large consensus today on the fact that this original name was very misleading, suggesting the existence a singularity with infinite density at zero time while this is based on extrapolating the Friedman-Lemaître-Roberston-Walker before the Planck time ($~10^{-43}\mathrm{s}$), therefore beyond the domain of validity of General Relativity. There is currently no convincing theory to describe such conditions where spacetime curvature and quantum physics need to be accounted for simultaneously and this primordial epoch is therefore not described in the current cosmological model. It is likely (although not proven) that a correct description of this era would show that the singularity is actually avoided (be it through strings, loop quantum gravity or other theories).

Before describing observations that allow to measure the actual value of various parameters of the $\Lambda$CDM model, it is important to investigate the observations that broadly point towards this model, or at least give arguments supporting the general idea of an expanding Universe that was denser and warmer in the past.

\subsubsection{Expansion}
The first of these arguments is of course the original one: the evidence for the Hubble law \citep{hubble} showing that distant galaxies experience a redshift proportional to their distance. The direct interpretation of this redshift as a Doppler effect resulting from the recession velocity of distant galaxies in the usual sense is slightly oversimplified but broadly accurate in the framework of General Relativity \citep{bunn}. The only sensible way of interpreting this apparent velocity from distant galaxies under the assumption of the Cosmological Principle is through a global expansion of the Universe. In such a global expansion picture, all observers see distant objects redshifting from them with the same Hubble law. After decades of longstanding debates on the actual value of the Hubble constant $H_0$, the situation now seems to be stabilized to $H_0=72\pm3\mathrm{(stat)}\pm7\mathrm{(syst)}$ \citep{freedman}. The main systematic effects limiting this measurement are related to intermediate distance calibrators such as the cepheids metallicity and the distance to the Large Magellanic Cloud. The observation that the Universe is expanding is at the origin of the idea by G. Lemaître that the galaxies were much closer to each other in the past (the primeval atom), hence corresponding to a denser Universe in the past, precisely the FLRW picture. After decades of measurements of astrophysical objects at larger and larger distances, their increasing redshift has suffered no exception confirming the initial explanation of an expanding Universe. The relation between measured redshifts and actual distances (the Hubble Law itself) is at the basis of most of the cosmological tests that check for consistency of the observations with the $\Lambda$CDM model and measure its parameters, including the explanation of the accelerated expansion through Dark Energy (see section~\ref{dark_energy}).

\subsubsection{Cosmic Microwave Background}
The definitive evidence for this hot and dense past was brought by the discovery in 1965 of the Cosmic Microwave Background (CMB) \citep{penzias_wilson} immediately interpreted by Dicke, Peebles, Roll and Wilkinson \citep{dicke} as the cosmological relic radiation predicted in 1948 as a consequence of the Big-Bang model by Gamow, Alpher and Herman \citep{gamow, gamow1, alpher_herman}. This radiation is extremely isotropic and is that of a black-body at a temperature of 2.728~K. It results from the decoupling of photons from matter that occurred when the Universe transited from ionized to neutral at a redshift of $\sim$1100.  The measurements of the spectrum of the CMB are now so accurate \citep{mather} that it is the most perfect blackbody ever measured in nature. The existence of the CMB is in itself a very strong support to the Big-Bang model in the sense that it confirms that the Universe was indeed so hot and dense in the past that it was ionized, its uniform temperature is the best argument for the isotropy of the Universe (as was already discussed above). The amplitude of the small temperature inhomogeneities observed points towards a Universe dominated by cold (heavy, non-relativistic particles) dark matter (see section~\ref{cmb} for a more detailed discussion). Very impressive measurements of the history of the CMB temperature were obtained through measuring the {\em thermal Sunyaev Zel'dovitch} \citep{sz1,sz2} in galaxy clusters at redshifts around 0.5 \citep{luzzi} and at further distances observing the rotational excitation of CO molecules due to their illumination by the CMB in the spectra of distant quasars out to $z=3$ \citep{noterdaeme}. All of these measurements show a perfect consistency with the expected $(1+z)$ scaling of the temperature of the CMB in the framework of FLRW cosmology.

\subsubsection{Big-Bang Nucleosynthesis and light elements abundances}
Another main observational argument favoring the Big-Bang paradigm is the excellent agreement between theoretical calculations of the abundances of the light elements and the actual measurements. The Big-Bang Nucleosynthesis theory was first developed by Alpher and Gamow in the late 40's \citep{alpher,gamow}, it is based on the nuclear cross sections between light nuclei and shows that the photons, protons and neutrons present in the early times start to form heavier nuclei while the temperature of the Universe cools down during the first minutes after the Big-Bang. These nuclear reactions are frozen when the temperature becomes too low and when neutrons start to decay massively because of their short lifetime. Basically, the only free parameter is the baryon density, measured by the parameter $\Omega_b$, the fraction of the volume density of baryons to the critical density of the Universe (the density for which the curvature is zero in FLRW models). The primordial abundances of the light elements can be measured in the Universe by observing regions that experienced little stellar nucleosynthesis such as metal poor stars or very distant objects such as in the Lyman-$\alpha$ forest of quasars. Although these measurements are extremely difficult as fusion occurring in stars modifies the picture in a non-negligible manner, the observations agree very well with theoretical predictions for a baryon fraction of $\Omega_b h^2\sim 0.02$ (where $h$ is the normalized Hubble constant $H_0/100\sim 0.7$, often used to quote cosmological quantities whose actual value is degenerated with the Hubble constant)  which is a very small amount of baryonic matter suggesting a very low density Universe or a Universe whose matter part is dominantly dark as stated by the $\Lambda$CDM model. This very impressive agreement has been further confirmed with an independent measurement of the baryon fraction from Cosmic Microwave Background observations with WMAP \citep{wmap7_komatsu}. There is however a discrepancy  in the Lithium-7 abundance measurements with predictions from the standard Big-Bang Nucleosynthesis model (it is observed around a factor 3 less than predicted). An unbiased measurement of the primordial Lithium-7 is particularly difficult as it could easily be burnt in stars. It is therefore not yet clear whether the problem lies in the nucleosynthesis theory or in the interpretations of the measurements \citep{iocco}.

\subsubsection{Reionization}
A more recent argument favoring the broad picture of the hot Big-Bang model is known as reionization. After the decoupling of the photons from the the baryons and the emission of the Cosmic Microwave Background (see section~\ref{cmb}), the Universe entered in a phase known as the {\em Dark Ages} where most of the structure formation occurred. The name {\em Dark Ages} comes from the fact that the only light propagating in the Universe at this epoch was the CMB that became more and more redshifted. While structure formation went on, regions of the Universe started to heat and emit small amounts of light at wavelengths smaller than the Lyman-$\alpha$ line that could not propagate freely as it was absorbed by neutral hydrogen (through ionization). It is only when the first stars and quasars formed (between 150 and 800 million years after the Big-Bang) that their UV light started to massively ionize the surrounding hydrogen clouds, allowing this high energy light to propagate through the Universe, percolating through larger and larger ionized regions. Eventually, at a redshift around 6, the Universe ended up being totally ionized, this is the moment known as reionization. 
Although we have to wait for the next generation of 21~cm line interferometers to fully explore the dark ages and reionization, we already have direct evidence of the validity of the above scenario through the observation of the Gunn-Peterson trough predicted decades before its observation \citep{gunnpeterson} in the spectrum of distant quasars. Basically, light from the most distant quasars, emitted before reionization was complete should be totally absorbed at wavelengths below the Lyman-$\alpha$ line as such photons would have been absorbed in ionizing the dominantly neutral hydrogen. Less distant quasars should not exhibit this trough as they emitted their light in an ionized Universe where these photons propagated freely except when they accidentally encountered rare regions of still neutral hydrogen (forming the Lyman-$\alpha$ forest). The Gunn-Peterson trough was first observed with a quasar at  redshift 6.28 \citep{becker} showing absolutely no emission at wavelengths below the Lyman-$\alpha$ line. Such an observation was later confirmed by others showing that reionization was complete below redshift 6.  Reionization also affects the CMB anisotropies observed by WMAP, producing a specific polarized feature on the large scales, allowing for an independent measurement of the reionization history. The latest results from WMAP \citep{wmap7_komatsu} indicate that reionization should have started around redshift 11 and ended around redshift 7. This result is in rather good agreement with the Gunn-Peterson trough data considering the large uncertainties in the WMAP measurement. The Planck satellite is expected to improve this measurement significantly in the next few years.
Similarly as in the above cases, the new observations match extremely well with the picture of the Big-Bang scenario in which reionization is a unique predicted feature.

All of the observations mentioned in this section point towards a Universe with a history: it appears to be expanding now, to have been expanding for a long time as we can trace back important events related to higher densities and temperatures: Gunn-Peterson trough related to reionization, CMB temperature measurements at various redshifts, the emission of the CMB itself corresponding to a temperature related to the hydrogen binding energy (modulated by the very high photon to baryon ratio) and Big-Bang Nucleosynthesis related to energies corresponding to nuclear reactions. In that sense, all of these observations strongly support the hot Big-Bang paradigm which not only predicts these important events but allows one to calculate observable quantities and obtain excellent agreement between calculations and observations.

\section{The Cosmic Microwave Background\label{cmb}}
The existence of the relic Cosmic Microwave Background was predicted by Gamow, Alpher and Herman in 1948 \citep{gamow, alpher_herman} as the consequence of the cooling of the primordial plasma. At early times, the high temperature of the Universe prevents electrons from being bound to nuclei due to the high energy of the photons interacting with them constantly. This results in a perfect thermalization of these species. Eventually, as the Universe expands, the photons loose energy and electrons start to be captured by nuclei. This happens at a significantly lower energy than the hydrogen binding energy (13.6 eV) because the photons outnumber the baryons by a factor of about one billion so that even a lower energy, the high energy tail of the Planck distribution of the photons still keeps the Universe ionized. However, at a temperature of around 3000K (0.25 eV), at a redshift $z~\sim 1100$, the Universe becomes neutral. This moment is known as the {\em matter-radiation} decoupling\footnote{There is actually an earlier helium decoupling \citep{sunyaev}: He nuclei recombining with one electron at $z\sim 7000$, followed by the recombination of the second electron at $z\sim2500$).}. This results in an increase of the mean-free-path of the photons beyond the Hubble radius so that the Universe becomes transparent to these photons\footnote{As said in the section dedicated to reionization, from this point the Universe is opaque to high energy photons that would be absorbed by neutral hydrogen through ionization. But the CMB photons are below this threshold and therefore do not interact with hydrogen anymore.}. They are released with a perfect blackbody spectrum at a temperature of 3000K and are observed today as the Cosmic Microwave Background radiation with a blackbody temperature of 2.7K, the emission peaking at wavelength around 2~mm. As mentioned before it was discovered by Penzias and Wilson in 1965 \citep{penzias_wilson} and immediately interpreted as the relic background from the hot past by Dicke, Peebles, Roll and Wilkinson \citep{dicke} according to the prediction from Gamow, Alpher and Herman \citep{gamow, gamow1, alpher_herman} in 1948.

\subsection{The corner stone of the Big-Bang theory}
Besides being a major observational evidence for the hot Big-Bang model (see above), the CMB is nowadays the cornerstone of the $\Lambda$CDM model thanks to exquisite observations of its temperature anisotropies achieved during the last 20 years. Apart from the discovery of the CMB itself, the success-story of the CMB started in 1992 with the NASA COBE satellite that embarked two instruments dedicated to CMB: FIRAS was designed to confirm the blackbody nature of the CMB and obtained such a good measurement of this spectrum \citep{mather} that the errors bars on the famous plot showing the spectrum needed to be multiplied by 400 to be noticeable. The second instrument, DMR, was designed to map the temperature of the CMB with an angular resolution of a few degrees and with high accuracy in order to find anisotropies in the CMB. From the beginning, such anisotropies were expected in the CMB due to the density fluctuations present in the Universe at early times. These were expected to be the seeds for the formation of structures that are observed today. At this time, the Hot Dark Matter scenario (where the dark matter particles are light and therefore relativistic, such as neutrinos) was the preferred one and required relatively large temperature anisotropies that remained undetected. This pushed theoreticians to move to the Cold Dark Matter scenario (with massive, non-relativistic dark matter particles, therefore implying new physics -- see section~\ref{dark_matter}) which proved successful when COBE/DMR found the anisotropies at the level expected from Cold Dark Matter: fluctuations at the $10^{-5}$ level, around 30~$\mu\mathrm{K}$ on the 2.7K average temperature of the CMB \citep{smoot}. 

\subsection{Baryonic acoustic oscillations}
The temperature fluctuations in the CMB are the consequence of the density fluctuations in the plasma before decoupling. These evolved from the primordial fluctuations in a way that relies on well known physics and can therefore be calculated accurately. This evolution is mainly due to baryonic acoustic oscillations and has the nice property of creating a very visible feature in the matter power spectrum at decoupling, and therefore in the CMB temperature fluctuations. When a primordial density fluctuation of a given size ``enters its horizon"\footnote{The reasoning here is done in Fourier space, meaning that the matter field is considered scale by scale instead of a function of its space--time coordinates. A fluctuation of a given scale ``becomes aware of its own gravity" (and is therefore not only affected by the background expansion but also by its own gravity and other physical processes) at a time given by its scale divided by the speed of light. We usually say that this perturbation enters its {\em horizon}, the horizon being the distance from a given point reachable by information traveling at the speed of light since the Big Bang.}, it may start to collapse falling into the potential well formed by dark matter that decoupled from the photon-baryon plasma at early times (and therefore started to collapse earlier). The gravitational collapse can however only efficiently occur when gravity forces exceed the pressure forces due to radiation and is therefore heavily suppressed in the early times when the Universe is still radiation dominated. It is only after matter/radiation equality ($z_\mathrm{equality}\sim 3400$) that structures smaller than the so-called Jeans scale (where gravity and pressure forces are equal) start to collapse. This locally increases the density and temperature, resulting in an increase of the radiation pressure that pushes matter outwards (acting as a repulsive spring), reducing again the density and temperature, allowing matter to collapse again and so on... This oscillatory process is known as the {\em baryonic acoustic oscillations}. The most important thing to bear in mind is that the oscillations only start for scales that are smaller than the horizon. Larger and larger scales therefore undergo this oscillatory process as times goes on, and all scales of a given size in the Universe are at the same phase of their oscillation at a given time. When matter-radiation decoupling occurs at a redshift of $z_\mathrm{decoupling}=1100$, the oscillations are brutally frozen as the photons escape, removing  their role of a spring in this oscillatory process. Matter then starts to really collapse and to form structures. But both the matter distribution and the photon bear the imprint of the oscillations that happen just before decoupling. Their power spectra show an oscillatory feature, the largest scale having oscillated being given by the sound horizon at the time of decoupling.

The baryonic acoustic oscillation pattern is observed with high significance as a series of peaks in the angular power spectrum of the Cosmic Microwave Background seen by WMAP \citep{wmap7_komatsu} and was first detected by ground based and balloon-born experiments in the early 2000s (Boomerang \citep{boomerang}, Maxima \citep{maxima}, Archeops \citep{archeops}). It was calculated theoretically in the framework of the Big-Bang model by Bond and Efstathiou in a seminal article in 1987 \citep{bond_efstathiou}, five years {\bf before} the discovery of the first anisotropies and 14 years before the clear detection of the first acoustic peak in the CMB. Since then, the WMAP data and other experiments provided exquisite measurements of the angular power spectrum (which will be further refined by the Planck satellite) showing an unprecedented agreement between a theoretical curve full of features, thousands of independent data points with a $\chi^2$ per degree of freedom amazingly close to 1. 

\subsection{Cosmological parameters}
The comparison of the observed CMB angular power spectrum with theoretical predictions for the primordial fluctuations power spectrum (usually calculated from inflationary models) allows to fit with high accuracy most of the $~12$ free parameters of the $\Lambda$CDM model\footnote{Note here that this fitting is done {\em within} the $\Lambda$CDM model and is essentially a check for consistency of the model. Besides the assumption of the validity of $\Lambda$CDM itself, some important assumptions are made here on fundamental quantities that do not explicitly appear as parameters: number of dimensions, fundamental constants, …} showing excellent agreement with other independent probes \citep{wmap7_komatsu}. It is particularly impressive to see the excellent agreement between the baryonic density obtained from the measurement of light elements in the Universe and Big-Bang Nucleosynthesis theory \citep{bbn_ob}: $\Omega_b h^2 = 0.0214 \pm 0.002$ and from CMB observations \citep{wmap7_komatsu} $\Omega_b h^2 = 0.02258 \pm 0.00057$ which are based on completely different physical processes: the former comes from the nuclear reactions in the three first minutes of the Universe while the second comes from the impact of the baryonic density on the Baryon Acoustic Oscillations happening 400 000 years after the Big-Bang. The measurement of the matter content, curvature of the Universe, dark energy density and hubble constant are also extremely consistent with other probes: the CMB itself hardly puts strong constraints in the $(\Omega_m,\Omega_\Lambda)$ plane due to a strong degeneracy with the value of the Hubble constant. However, if one assumes a value for the Hubble constant as measured by the HST key project of $H_0\sim72\mathrm{km.s^{-1}.Mpc^{-1}}$ the allowed values from CMB+H$_0$ values for $\Omega_m$ and $\Omega_\Lambda$ reduces considerably and points towards a flat Universe\footnote{The results show consistency with flatness within some observational uncertainties. This means that we do not know the sign of the actual curvature and that the fate of the Universe (recontraction or not) is still unknown.} with $\Omega_m+\Omega_\Lambda\sim1$ where $\Omega_m\sim0.3$ and $\Omega_\Lambda\sim 0.7$, very consistent with other probes such as type Ia supernovae and observations of Baryonic Acoustic Oscillations in the distribution of galaxies.

\subsection{Polarization}
Besides temperature anisotropies: the anisotropic flow of photons on electrons around density perturbations just before matter-radiation decoupling induces partial polarization of the initially unpolarized radiation. As a result, the CMB is polarized at the $\sim 10\%$ level and also exhibits polarization fluctuations at the $\mu\mathrm{K}$ level. These can be of two types, the scalar E-polarization, or the pseudo-scalar B-polarization. Four angular power spectra can then be extracted from the full cross-correlation of the CMB maps (temperature and polarization): $C_\ell^\mathrm{TT}$, $C_\ell^\mathrm{TE}$, $C_\ell^\mathrm{EE}$ and $C_\ell^\mathrm{BB}$, the two remaining ones being zero by symmetry. E polarization was detected for the first time in 2001 by two interferometers DASI \citep{dasi} and CBI \citep{cbi}. The first three of the above spectra are now observed with excellent accuracy with WMAP \citep{wmap7_komatsu}  and will be ultimately measured by the Planck satellite. As mentioned before, the most favored model for generating primordial perturbations is {\em inflation}, a period of accelerated expansion thought to have occurred around $\sim 10^{-35}\mathrm{s}$ after the Big-Bang. Inflation, on top of solving known issues in the Big-Bang model (curvature, horizon and monopoles) offers a very attractive mechanism for producing scalar perturbations of the metric that are in impressive agreement with the observations\footnote{It is important to remark however that the detailed mechanism allowing the transition from quantum to classical fluctuations during inflation is still not fully understood \citep{perez,landau}.}. In particular, in the simplest inflation models, the perturbations are expected to be adiabatic (all the species fluctuate in phase) which results in a very specific property of the CMB polarized power spectra: the series of peaks (coming from baryon acoustic oscillations) in the temperature power spectrum should be shifted by half a period with respect to that of the E-mode polarization.  Peaks in $C_\ell^\mathrm{TT}$ should correspond to troughs in  $C_\ell^\mathrm{EE}$ which is indeed observed (again after the prediction) while it is not a trivial property. In addition to this property, the general shape of the observed power spectra is in agreement with the predictions of inflation. In particular the spectral index of the density perturbation, measured to be $n_s=0.963\pm 0.014$ by WMAP \citep{wmap7_komatsu}, is expected to be slightly below 1 in inflationary theories\footnote{The spectral index tends to one after inflation and only reaches it for an infinite inflation.}.

\subsection{The quest for B-modes}
The last of the four angular power spectra observable from the CMB, the B-modes one $C_\ell^\mathrm{BB}$ still awaits detection. B polarization can only be induced by tensor perturbations of the metric, a specific prediction of inflation. These perturbations correspond to primordial gravitational waves produced at the end of inflation. Their amplitude relative to the density perturbations is the {\em tensor-to-scalar ratio} $r$, whose measurement (along with the tensor spectral index $n_t$) would allow one to discriminate among the large number of inflationary models that currently agree with the observations. The parameter $r$ is expected to be small (simple inflationary models are in the range $10^{-3}<r<0.1$) so that the corresponding B-polarization is expected to be very small, a few $10~\mathrm{nK}$ and therefore extremely difficult to detect because of foreground emissions (from dust or synchrotron emission in the galaxy) and instrumental depolarization that mixes E and B (the former being much larger than the latter). A number of teams around the world have however undertaken this quest for the B-mode polarization and hope to be able to detect it within the next decade. Such a detection is considered by the community as the smoking gun for inflation because tensor perturbations can only be created in this framework (at least up to now). The actual value of $r$ and, although it will be even harder to measure, the value of $n_t$ would allow one to understand the inflationary process in its details such as constraining its energy scale and the shape of its potential \citep{baumann}.

\subsection{CMB achievements}
Considering the number of major achievements in our understanding of cosmology and in the consolidation of the $\Lambda$CDM model obtained thanks to the CMB measurements, it is worthwhile to summarize them:
\begin{itemize}
\item {\bf 1965: discovery of the CMB by Penzias and Wilson}. The CMB appeared as an isotropic black-body radiation as predicted by Gamow, Alpher and Herman in 1948 within the hot Big-Bang model. This was the time of the crystallization of the community around the Big-Bang theory.
\item {\bf 1992: COBE satellite}. While FIRAS confirmed the black-body nature of the CMB, DMR discovered the first anisotropies at the level of $10^{-5}$. This favored the Cold Dark Matter model to account for the dark mass massively present in the Universe and responsible for structure formation.
\item {\bf 1999: Boomerang and Maxima}. These two balloon borne experiments measured the anisotropies down to sub-degree scales showing the baryonic acoustic oscillations peak at an angular scale corresponding to a flat Universe. The shape of the peak also excluded the topological defects from being responsible for the seeds for structure formation and therefore strongly favored inflation. The shape of the angular power spectrum proved to be excellent agreement with predictions made 15 years before by Bond and Efstathiou.
\item {\bf 2001: DASI and CBI}. These two interferometers detected polarization of the CMB for the first time. It had also been predicted by Bond and Efstathiou. Fitting the power spectra on temperature anisotropy data and comparing the expected polarization spectra to the measurements showed excellent agreement, a major confirmation of the coherence of the model. 
\item {\bf 2003: WMAP}. This NASA satellite provided exquisite full-sky measurement of the CMB anisotropies in both temperature and polarization down to $\sim 20$ arcmin resolution. This was the revolutionary dataset that allowed establishing the $\Lambda$CDM model and confirming a complete coherence between the model and the observations.
\end{itemize}
The next step is expected to be achieved by the ESA Planck satellite that will improve further the measurements obtained with WMAP and allow to reduce the uncertainties on cosmological parameters by a factor three with higher sensitivity and angular resolution. After Planck, the whole CMB community is turning towards the B-mode polarization quest, the next big step in observational cosmology. Ground based, balloon-borne or satellite experiments are currently planned to tackle down B polarization in the next decade and start exploring the inflationary era.

\section{Dark Matter\label{dark_matter}}
As said before, the composition of the Universe measured by many independent probes is very puzzling: matter seems to account for barely more than one fourth of the energy content of the Universe and most of this matter is identified as being dark, as opposed to only 4 percent in the form of ordinary matter (stars and gas clouds). Evidence for this dark component in the matter budget of the Universe comes from various probes that show that the motions of luminous objects is not consistent with what one would expect from them alone. The motions are better explained if one assumes that a large fraction of the mass distribution is dark with a similar repartition to that of luminous matter although more diffuse on the small scales (note that the relation between luminous and dark matter, known as the {\em bias} is still largely debated). By dark, one of course assumes that it does not produce light, but in a broader sense that it is interacts in a very week manner with ordinary matter (hence the absence of light) and with itself. It is therefore expected to be non-collisional.
It appears that the larger the scales considered, the more dark matter is required to explain the observations. An intensive observational effort has been undertaken in the last 30 years to identify this dark matter without any success up to now. However, dark matter is needed in such a consistent manner in the standard cosmological model (the CDM in $\Lambda$CDM stands for Cold Dark Matter) that most cosmologists have little doubt of its existence. It is useful to remark \citep{aubourgHDR} that from the methodological point of view, inferring the existence of unobserved matter from its gravitational effects proved successful in the past: Le Verrier predicted the existence of the planet Neptune from the observed motion of Uranus. At the same time, Bessel predicted that Sirius was a binary star from the observed motion of it bright component. New extrasolar planets are now routinely discovered using the same technique. Note that, of course, this reasoning does not work systematically, the best example being the prediction by the same Le Verrier of another planet, Vulcain, to account for the anomalous precession of the Mercury's perihelion within Newtonian gravitation. The correct explanation turned out to be to change gravitation theory to General Relativity... Therefore, another way of solving the problem is to assume that the laws of gravitation that one applies are not correct and that a modified theory of gravity could account for the observations without requiring a dark component. This approach is well motivated in the sense that dark matter is precisely seen only through unexpected gravitational effects. Although General Relativity is well tested in many circumstances, it would be too daring to consider it  as the final words on gravitation.
Such efforts to explain dark matter through modified gravity are also undertaken both from the observational and theoretical sides but did not yet shed much light of this dark mystery~\citep{martineztrimble}.

\subsection{Evidence for Dark Matter}
The first evidence for dark matter is usually attributed to Fritz Zwicky in 1933 but, as quoted in \citep{aubourgHDR}, earlier articles mentioned the possibility that a significant amount of dark matter may exist: \"Opik in 1914 and Kapteyn in 1922 cite the words "dark matter" for the first times \citep{opik,kapteyn} and suggest it could be measured from the motion of stars in the vicinity of the Earth. Jeans concludes from a similar study in 1922 that observations suggest three times as much dark matter as visible matter in the Universe \citep{jeans}. In 1932, Oort improves these observations and draws similar conclusions \citep{oort}. Actually, the need for dark matter to explain the motion of nearby stars is nowadays questioned and the problem could be inverted in the sense that these motion are found by some authors to be inconsistent with the average amount of dark matter expected locally~\cite{bidin}. This recent work has however been severely criticized~\cite{bovy} and should be considered with caution.

\subsubsection{Cluster dynamics}
In 1933, Zwicky tried to explore the Hubble law by further extragalactic observations and studied the dynamics of the galaxies in the nearby Coma cluster \citep{zwicky}. From the velocity dispersions of the galaxies he calculated the kinetic energy in the cluster, and from the distances between galaxies he obtained the potential energy. Applying the Virial theorem, he was able to infer that the total "gravitational mass" of the Coma cluster was from 100 to 500 times larger that of the galaxies. He concluded that such a "dark matter" needed to be massively present in galaxy clusters. It is interesting to note that this article was cited very rarely until the 80s when the community realized that dark matter had a very strong observational basis. Cluster dynamics could be explained through dark matter, be it baryonic or not, or by a change in the theory of gravitation at the required scales.

\subsubsection{Rotation curves}
This resurgence of interest to dark matter was triggered by the observations made by V. Rubin in the 70s of the rotation speed of stars in the spiral galaxies, summarized in a famous article in 1980 \citep{rubin}. These showed a basically constant speed way beyond the regions of high stellar density, showing that the mass density needed to be constant far from the luminous center of the galaxies, requiring a vast halo of dark matter extending beyond luminous stars.  From this moment, many other observations came to confirm this work in various systems. It is important to note here that such rotation curves could be explained by a halo of dark (baryonic or not) matter but may also be explained by modifying gravitation as proposed within the MOND paradigm which will be discussed later.

\subsubsection{Hot gas in clusters}
The birth of X-ray astronomy in the early 70s revealed intense X-ray emission from galaxy clusters that was soon attributed to the presence of very hot gas in these clusters extending far beyond the galaxies \citep{kellogg}. Although there was some hope to explain the cluster dynamics with this gas (therefore baryonic "almost dark" matter), it appeared immediately that, based on the assumption of hydrostatic equilibrium, the gas could only account for half the mass of the cluster and that to explain the high temperature of the gas ($10^7-10^8~\mathrm{K}$) at least 85\% more mass was needed, requiring, like in Zwicky's work, large amounts of dark matter (baryonic or not).

\subsubsection{Gravitational lensing}
It was also Zwicky who first proposed in 1937 to measure the mass of galaxies through the deflection of light predicted from General Relativity \citep{zwicky2}. This gravitational lensing effect is expected when a massive foreground object (such as a cluster) deflects the light form background objects (field galaxies). This technique experienced amazing developments in the last decade revealing strong distortions when there is an alignment between the lens and the background objects, and weak lensing for the off-axis field galaxies. The latter technique allows to map in a very accurate manner the distribution of gravitational matter in the foreground clusters. Such studies performed on a number of clusters clearly showed that gravitational matter extends beyond the galaxies, giving further evidence for dark matter. The most well known example of such a study is given with the "Bullet Cluster" \citep{clowe} where the distribution of matter is mapped using the three possible probes: galaxies through their visible emission, hot gas through X-ray observations (by the Chandra Satellite) and gravitational matter through weak lensing. The bullet cluster is actually formed by two interacting cluster, one of them has gone through the other recently (hence the name "bullet"). On the images superimposing the three probes, a shock wave can be clearly seen in the gas where the collision occurred, and where the collisional gas was stopped by the shock. The galaxies have not experienced anything during the collision and pursued their trajectories normally. As expected from dark matter scenarios, the gravitational matter did not experience the shock either and remained with the galaxies leaving the gas behind. This is considered by many as the best direct evidence for dark matter and is indeed very impressive as one really needs massive, dark and non-collisional (therefore non-baryonic) dark matter to explain the observations. One should however note that another cluster seems to be a counter-example. Abell 520 is also a system in which two clusters have collided but where the gravitational matter cores apparently coincide with the gas rather than with the galaxies \citep{mahdavi, jee}. There are arguments to explain this observation by other means, such as an unfortunate alignment of a dark matter filament with the line of sight towards the cluster. The situation is in any case still unclear and one should be cautious in drawing too strong conclusions on the basis of few of these cluster collisions systems. In the next few years, more objects of this kind will allow one to conclude in a more definitive manner.

\subsubsection{Structure formation}
As mentioned above (section~\ref{cmb}), the structures we observe around us would need large ($\sim10^{-3}$) density fluctuations at the time of the decoupling between photons and baryons to explain them. The CMB fluctuations are observed two orders of magnitude below requiring matter to have started to collapse earlier than at the time of decoupling and to have more mass to counterbalance locally the global expansion and form structures. Dark matter comes here again to solve the problem: if dark matter is dark it is because it is decoupled from ordinary matter. If this decoupling has occurred before the matter-radiation decoupling, then clusters of dark matter already collapsed significantly before the emission of the CMB creating potential well large enough to explain structure formation with the small ordinary matter perturbations observed in the CMB. Further to being decoupled before recombination (and therefore non-baryonic), the dark matter particles also need to be massive enough not to be relativistic, otherwise their free streaming would erase density fluctuations on small scales which is not compatible with the observations and magnitude of the CMB fluctuations. Such a non-relativistic dark matter is known as Cold Dark Matter and is a key ingredient in the $\Lambda$CDM scenario.

\subsubsection{Baryonic Acoustic Oscillations\label{dmbao}}
The baryonic acoustic oscillations were already discussed in the CMB section (section~\ref{cmb}) but are also the most convincing evidence for the existence of dark matter (in the author's opinion). Further to being responsible for the series of peaks observed in the angular power spectrum of the CMB \citep{wmap7_komatsu}, they are also observed in the distribution of galaxies as a peak in their two-point correlation function. The BAO can be seen as the propagation of a pressure wave in the matter-radiation plasma that travels at the speed of sound from the center of each perturbation where the dark matter stays and collapses. When matter and radiation decouple, the sound speed drops to zero and the wave stalls. The photons, dragging baryons, escape and there remains an excess density with the form of a sphere located at the sound horizon (roughly 150 Mpc in comoving distance) from the center of the perturbation. This spherical excess of matter has an excess of dark matter at its center. Both excesses tend to fall into each other and equalize their density contrast but a very specific pattern remains: an excess of matter and dark matter at the center of each initial perturbation surrounded by an excess on a sphere at 150 Mpc. Galaxies later form preferentially in these regions imprinting a peak at a scale of 150 Mpc in the two-point correlation function of large galaxy catalogs (it is also seen equivalently as wiggles in the Fourier power spectrum of these catalogs). Note that this 150 Mpc scale was predicted from first elements a long time before its actual discovery.
The discovery was done almost at the same time with the SDSS data (luminous red galaxies) at a redshift of 0.35 \citep{eisenstein} and the 2dFDRS data \citep{cole2dF}. It was followed by other observations with the 6DF survey at z=0.1 \citep{beutler}, the Wiggle-z survey at z=0.6 \citep{blake} and more recently by the BOSS survey \citep{anderson}.
This BAO feature is used as a standard ruler to put constraints on dark energy but is also the best evidence for dark matter. If there was no dark matter, there would only be an excess on a sphere at 150 Mpc and nothing at the center. This would not result at all in the same feature in the two-point correlation function, but on a broad excess with a drop at twice the sound horizon ($\sim 300$~Mpc) which is rejected by observations. The clear peak, detected at the five standard deviations level in the BOSS data \citep{anderson} is therefore a strong evidence for the presence of a massive component that was decoupled from the plasma earlier than recombination as it stayed behind when the sound wave propagated. This component is therefore expected to have very suppressed interaction with ordinary matter, and particularly with photons and therefore to be dark. This is the exact definition of non-baryonic dark matter. It seems very difficult to mimic such a feature in the correlation function of the galaxies, and at least no such scenario was ever proposed with concurrent models to dark matter (MOND for instance).

\subsubsection{Big-Bang Nucleosynthesis}
Big-Bang Nucleosynthesis results were already mentioned above, they bring tight constraints on the amount of baryonic matter in the Universe, pointing towards a very low baryonic matter content: $\Omega_b h^2\sim 0.02$. Such a low value is incompatible with all other measurements of the dynamics of structures in the Universe (showing a total density of $\Omega\sim0.3$) unless most of the matter is in the form of non-baryonic dark matter.

\subsection{Dark matter scenarios}
Faced with the major issue of explaining how a major fraction of the matter of the Universe could be in the form of dark matter, theorists have proposed various scenarios. Most of these are based on speculative new physics as there is no simple way to explain dark matter with well known models.
\subsubsection{Baryonic Dark Matter: gas and Machos}
Although the largest part of dark matter needs to be non-baryonic (as it needs to be weakly interacting, decoupled prior to baryons), there is also a fair amount of missing baryons in the Universe. These baryons could be in the form of cold gas characterized by low emission for which there is little observational constraints. 

Another model was proposed in the 80s: MAssive Compact Halo Objects (Machos). These are stars with such a low mass (below 0.08 solar masses) that their temperature is not high enough to initiate nuclear fusion. They are therefore low emitters and could hardly be detected directly. They were considered as appealing as they could explain the galactic rotation curves without invoking new physics (gravity modifications or exotic particles). The only way to observe such objects appeared to be through the gravitational lensing they would cause when passing exactly in front of a background star in nearby galaxies. Such a lensing effect is expected to be extremely small, hence called microlensing, but would be observable in the case of a perfect alignment. The expected signature is a wavelength independent amplification of the luminosity of the background star on a few days timescale. Such event were searched for in the 90s by using the Magellanic Clouds and Andromeda Galaxy as a set of background stars. Although many such events were observed \citep{alcock, aubourg_eros}, confirming the viability of the detection technique, their number remained way too low to validate Machos as the explanation for the missing baryonic dark matter \citep{eros, agape, crotts}.

\subsubsection{Wimps and Axions}
The archetype of Dark Matter is a non-baryonic massive particle that has little interaction with ordinary (baryonic) matter.
This is precisely the definitions of WIMPs (Weakly Interacting Massive Particles). The existence of such a particle would satisfy all the observational features of Dark Matter and be the best candidate for the standard model's Cold Dark Matter. Such a particle doesn't exist in the Particle Physics standard model and therefore requires new physics. Supersymmetry is a very popular extension to the standard model and relies on a specific new symmetry between baryons and fermions that brings a whole new set of particles appearing at high energy (beyond a few TeV) that are expected to have low interactions with baryons. The lightest of these supersymmetric particles (often known as the {\em neutralino}) is expected to be stable (as it cannot decay into baryons) and to remain in our low energy Universe. These neutralinos are the best candidates for dark matter particle. 

Supersymmetry should exhibit visible signatures in high energy colliders such as the LHC. Unfortunately, there is no hint of the presence of such particles in the data collected so far \citep{olive}. The LHC results actually already put very tight constraints on the minimal supersymmetric models disfavoring them strongly. There are however non-minimal models with larger parameter space that are more difficult to constrain experimentally.

A more model-independent approach to WIMP direct detection is through their elastic scattering on nuclei in a target. Such a rare interaction could be detected by three different means, two of them being usually combined in a detector: production of heat, ionization and scintillation. The expected event rate is so low that a large target mass is required while the background needs to be reduced dramatically by insulating the detector from undesirable muons produced in the atmosphere that would mimic the dark matter interaction. The detectors are therefore buried under kilometers of rock in underground mines or tunnels. The CDMS \citep{cdms}, Edelweiss \citep{edelweiss} and Xenon \citep{xenon} experiments have excluded large regions of the WIMP mass versus cross-section plane (with best sensitivity for a cross-section of $10^{-43}-10^{-44}~\mathrm{cm^2}$ for 50 GeV WIMP mass) from the absence of such interactions in their data. Other experiments, such as DAMA/LIBRA \citep{dama} have observed an annual modulation of their event rate (without background subtraction) that could be attributed to the varying speed of the Earth in the dark matter halo of our Galaxy. The signal reported by DAMA/LIBRA is however inconsistent with the excluded values from CDMS, Edelweiss and Xenon and is strongly debated in the community. Although a number of possible explanations have been proposed, the situation is still confused and no definite conclusions can be drawn from these direct searches for WIMPs. The next few years will see these experiments use more massive targets and will therefore allow for more precise measurements and hopefully a clarification of the experimental situation.

Indirect detection of dark matter has also been proposed through the annihilation of the WIMPs in the galaxy centers producing high energy photons that could be detected by $\gamma$-ray observatories such as HESS, CTA or the Fermi satellite. Such an annihilation would also produce neutrinos and antimatter and are therefore searched for by experiments sensitive to these particles. No evidence for a signal from the galactic center has been found by neutrino or $\gamma$-ray experiments, but PAMELA \citep{pamela} has reported a significant excess in the positron flux that could be explained by dark-matter annihilation. Although a very interesting and popular possibility, indirect detection of dark matter relies on strong assumptions and fine tuning for the WIMPs and can be mimicked by several classes of hard to understand violent astrophysical objects \citep{barger}.

Another kind of speculative weakly interacting particles is the axion. These particles were proposed to solve the problem of the CP violation in QCD and their possible existence, although not confirmed, is considered as very serious by most particle physicists.
Bearing extremely low mass, the axions could have been produced in large quantities in the primordial Universe and would be non-relativistic today so that they could be a nice solution to Cold Dark Matter. Dedicated experiments are currently running in order to find evidence for the presence of axions but are still unsuccessful.

\subsubsection{Modified gravity}
Modified Newtonian Dynamics \citep{mond, mondreview} is a very simple yet impressive idea proposed to explain the flat rotation curves of stars and HI regions in galaxies without relying on dark matter. The idea is that in the very weak field limit, the fundamental relation between the force applied to an object (here gravitation) and its acceleration is slightly modified. With only one extra-parameter (the pivot acceleration) flat rotation curves are obtained in a straightforward manner. The well known yet unexplained Tully-Fisher relation \citep{tullyfisher} between galaxies intrinsic luminosities and theirs stars rotation speeds is also explained in the same manner. Although very impressive at the galactic scales, MOND fails to explain the dynamics on galaxy cluster\footnote{unless adding a large amount of unobserved neutrinos, which finally reduces to adding some missing dark matter.} scales and is unable to explain the baryonic acoustic oscillations observed in the CMB and in the distribution of galaxies nor the gravitational lensing observed in galaxy clusters. As a result, despite an obvious success on the galactic scales, MOND does not appear as a solution to the dark-matter mystery on the large cosmological scales \citep{mondreview}.
More generally, as dark matter only manifests itself through differences between the observed expected gravitational effect of ordinary matter, it seems natural to try to find an explanation in modification of the gravity itself as explored by \cite{moffat}. Such a possibility is not ruled out by observations and can be seen as a serious alternative to the $\Lambda$CDM model. Gravity, on the other hand is tested with growing precision and still shows a perfect agreement with General Relativity~\cite{reid}.

\subsection{Dark Matter summary}
The ghost of dark matter has flown for almost a century above cosmology and astrophysics but the situation is still completely unclear. The more data accumulated and the more theories elaborated to explain the dark matter mystery, the darker the situation... On the small scales, the successes of MOND in explaining galactic dynamics tend to favor the idea of modified gravity while its failure on larger scales and the exquisite agreement between CDM scenarios and cosmological data strongly supports the idea of weakly interacting massive particles. The existence of such dark matter particles appears as the most sensible explanation for all of the observations described in this section. On the one hand, the evidence is clear for the need of massive particles, decoupled prior to the matter-radiation decoupling. On the other hand, the lack of success of the searches performed up to now revives the interest for other possibilities, such as modifications of gravity. We can only be optimistic and hope that the next few years, during which many experiments will have reached maturity, will provide the community with new, possibly decisive, informations regarding the nature of dark matter.

\section{Dark Energy\label{dark_energy}}
In the $\Lambda$CDM model, 70\% of the energy content of the Universe is modeled as an unknown form of energy, stretching space and triggering accelerated expansion. The data is consistent with a cosmological constant, a term in the Einstein equations that can be thought as a property of gravity or as the quantum field theory vacuum (a minimal energy density contained in vacuum arising from quantum fluctuations). More complex explanations involve a dynamic dark energy. Models differ through their equation of state (the ratio between pressure and density) that is constant $w=-1$ for a cosmological constant and different and variable with time for general dark energy models.
The situation is for now extremely unclear from the theoretical point of view, while the observations supporting evidence for Dark Energy are more and more diverse, convincing and consistent with a cosmological constant \citep{anderson, sanchez}.

\subsection{Evidence for Dark Energy}
The discovery of the accelerated expansion of the Universe through observing distant type Ia supernovae to be dimmer than expected \cite{riess, perlmutter} is usually considered as the first convincing evidence for Dark Energy. It is clear that the cosmological community experienced a dramatic transition when these results were released in 1998. There were however earlier claims for the possibility that a cosmological constant played a significant role in cosmology. Such a constant had originally been proposed by Einstein himself to preserve a static Universe within his dynamic equations for General Relativity. As was shown by Lemaître and Friedman, the presence of this cosmological constant was not sufficient to keep the Universe static and anyway the discovery of the expansion by Hubble \citep{hubble} removed the need for such a constant. When large scale structure observations started to be refined enough in the 80s to allow drawing  cosmological conclusions, it appeared that more structure was observed  than expected in a flat $\Omega_m=1$ model, suggesting that structure formation recently slowed down, which could be explained by the recent domination by a cosmological constant \citep{efstathiou}. Such considerations revived the interest for the cosmological constant. In the mid-90s, the observations of stars in globular clusters older than the estimated age of the Universe \cite{chaboyer} brought more arguments for the need for a cosmological constant that allows for an older Universe than calculated with matter-only (13.7 Gyr in $\Lambda$CDM as compared with 9.2 Gyr in a flat matter-dominated Universe).
When the type Ia supernovae distance measurements showed evidence for acceleration of the expansion, it appeared that the cosmological constant was the simplest explanation as it involved a single free parameter. Later evidence for the cosmological constant came to confirm this result so that Dark Energy (or in its simplest form, cosmological constant) seems very convincingly there from an observational point of view for most cosmologist while they easily recognize the lack of convincing theoretical explanation for it. Such an annoying situation justifies the tremendous efforts undertaken in the last decade and in the next one to tackle down dark energy using improved observations with diverse observational probes~\cite{weinberg}.

\subsubsection{Type Ia supernovae}
Type Ia supernovae are exploding stars originating from binary systems where a compact white dwarf captures material from its low density companion eventually reaching the Chandrasekhar mass where its supporting quantum pressure breaks down\footnote{A white dwarf is the last state of evolution of stars not massive enough to become neutron stars. All their hydrogen has been burnt in earlier stages so that only heavier nuclei remain in the white dwarf. There are no longer nuclear reactions in the white dwarf to support gravitational collapse, its stability is provided by electron degeneracy pressure, or quantum pressure that originates in the Pauli exclusion principle preventing two electron to be in the same quantum state. The density is so high in such a white dwarf that the exclusion principle acts as a pressure preventing the star to collapse and become denser. The quantum pressure is not infinite and is overrided by gravitational collapse when the mass of the star is larger than the Chandrasekhar mass of 1.44 Solar masses.}. The white dwarf then undergoes thermonuclear explosion sending heavy elements in the interstellar medium. A type Ia supernova is so violent that is intrinsic luminosity is of the order of that of a whole galaxy, allowing it to be observed from cosmological distances. Type Ia supernovae were shown to have an extremely regular absolute luminosity (within $\sim$15\%) \cite{phillips}, making them the ideal "standard candle" to measure distances throughout the Universe. The teams led by A. Riess and S. Perlmutter regularly observed patches of the sky in order to discover distant supernovae and use them to measure the cosmological parameters. In 1997, they obtained independently sets of measurements that allowed them to recognize that the supernovae at a given redshift were systematically less luminous than predicted by a Universe containing only matter \cite{riess, perlmutter}. This apparent faintness of distant supernovae could be explained if the distance to these supernovae was larger than expected. This could be explained in a standard FLRW scenario by the presence of a dominant (70\%) cosmological constant $\Lambda$ that triggered accelerated expansion. The fit of the FLRW model with $\Lambda$ to their data immediately appeared as very convincing. Three other possible explanations were however discussed: 
\begin{itemize}
\item {\bf Evolution:} The distant supernovae may not be exactly similar as the nearby ones used for calibrating their intrinsic luminosities because of galaxy evolution. They exploded in a much earlier phase of the Universe when galaxies where significantly younger. Further observations involving detailed spectra of nearby and distant supernovae taken at the same phase of their explosions showed a perfect matching of features in the spectra excluding significant effects from evolution.
\item {\bf Grey dust: } A possible low dense gray dust uniformly filling space could absorb the supernovae's light independently of the wavelength, mimicking the effect of a cosmological constant. In such a case, even more distant supernovae should be further dimmed which was not observed in subsequent datasets while it is naturally expected from $\Lambda$ that the acceleration effect should flatten beyond redshift $\sim 0.5$ due to the sub-dominance of the cosmological constant at higher redshifts.
\item{\bf Inhomogeneous models:} If we were located near the center of a large void, distant objects would be attracted by the walls of such a void and would experience accelerated recession from us. As discussed in section~\ref{cosmological_principle}, this possibility is now essentially ruled out although some debate still remains on this question. For instance, \citep{colin} claim that bulk flows from the inhomogeneous local Universe may affect significantly the Hubble diagram measurement at high redshift and reduce the significance of the evidence for acceleration with SNIa data.
\end{itemize}
Once these possible systematic effects were ruled out, the result was hard to criticize: the Universe was apparently experiencing accelerated expansion caused by an unknown 70\% dark energy. 

\subsubsection{Baryonic acoustic oscillations}
Once again, baryonic acoustic oscillations appear as an excellent tool for testing dark energy. As explained in section~\ref{dmbao}, the BAO imprints a characteristic scale of 150 Mpc in the matter distribution in the Universe. This scale can be used as a standard ruler in order to measure distances throughout the Universe. The BAO was detected at various redshifts allowing now to build a Hubble diagram similar to that obtained with luminosity distances using supernovae \citep{anderson, busca}. The resulting constraints strongly support the $\Lambda$CDM model. The latest BAO measurement to date, obtained at a redshift of 2.3 with the Lyman-$\alpha$ forest of distant quasars is the first precise measurement of the expansion rate in the decelerated phase (before dark energy domination). This BAO measurement also constrains the dark-energy density to be non-zero almost by itslef (only relying on the baryon density estimated by the CMB, but on no other cosmological emasurements). Further studies with BOSS~\cite{BOSS} will allow to strongly constrain the Dark Energy equation of state and hopefully help solving this mystery in the next few years.

\subsubsection{Concordance model and additional probes}
As was explained in section~\ref{cmb}, the Cosmic Microwave Background strongly constrains the geometry of the Universe to be flat (with $\Omega_m\sim 0.3$ and $\Omega_\Lambda\sim 0.7$) when combined with measurements of the Hubble constant \citep{wmap7_komatsu}. However, by itself, the CMB only brings a rather loose constraint in the $(\Omega_m,\Omega_\Lambda)$ plane in FLRW cosmology. The supernovae data constrains this plane in a completely different manner so that when combining CMB and supernovae (without the need for the Hubble constant), the allowed regions intersect on a flat Universe at the same place as the combination CMB+Hubble constant: $\Omega_m\sim 0.3$ and $\Omega_\Lambda\sim 0.7$. 

Furthermore, baryonic acoustic oscillations observed with the two-point correlation function from galaxy catalogues with SDSS \citep{eisenstein} and more recently with the BOSS data \citep{anderson,sanchez} independently constrain $\Omega_m\sim0.3$ supporting further this concordance model \citep{kowalski}. 

Latest CMB observations involving very small scales maps allow to measure the effect of the lensing by large scale structure on the CMB. This strongly depends on the value of dark energy and allows to have evidence for Dark Energy from the CMB itself combining data from WMAP and the Atacama Cosmology Telescope\citep{sherwin} without relying on external measurements of $H_0$ or $\Omega_m$. The central value obtained from this study is perfectly consistent with $\Lambda$CDM.

An additional probe for the presence of Dark Energy is the Integrated Sachs-Wolfe effect. The CMB photons are blue-shifted when falling into clusters potential wells and then red-shifted when escaping from these wells. Without Dark Energy, the net difference is zero but not in a Dark Energy dominated Universe. This effect can be seen as a correlation between the CMB temperature and the foreground galaxy distribution. Cold spots in the CMB are expected to correlate with voids in the galaxy distribution while hot spots correlate with massive clusters. Various studies have obtained evidence for this effect at the four standard deviation level~\citep{granett} giving another completely independent probe of the presence of Dark Energy. The values obtained with such an approach for the Dark Energy density are consistent with $\Omega_\Lambda\sim 0.7$.

Such a perfect agreement with all these independent probes giving consistent constraints on these two parameters, and supporting the theoretically expected flat model, is extremely impressive and gives a very strong support to the coherence of the data with the FLRW cosmology in general, and with the $\Lambda$CDM model in particular. As of today all measurements of the Dark Energy equation of state are consistent with a constant $w=-1$ corresponding to a cosmological constant although the currently modest accuracy of the measurement leaves space to a lot of alternative models.

\subsection{What could Dark Energy be ?}
The idea of a cosmological constant haunted observational cosmology since the very beginning (see \citep{rugh_zinkernagel} for a precise historical review): initially introduced by Einstein, disappeared for more than fifty years, brought back in the 1990s by the successes of the CDM model, along with the theoretical preference for a spatially flat Universe, while observations where showing more large scale structure than predicted with $\Omega_m=1$. This suggested a recent slow down of the structure formation that could be explained by the recent domination by a cosmological constant \citep{efstathiou}. Until direct evidence for an accelerated expansion came from the supernovae, the only model considered was the simple cosmological constant: a constant term in the Einstein equations making them completely general. But in the late 1990s, the standard model of particle physics was firmly established and a constant energy filling space such as the cosmological constant found a natural interpretation in this framework: vacuum energy, the zero-point energy of the vacuum state. However, in particle physics, vacuum energy is expected to be $10^{120}$ times larger than the value required to explain cosmological data\footnote{This is known as the worst theoretical prediction in the history of physics...}. In order to relate these two quantities, one would therefore need a mechanism involving a large but not total compensation which seems hard to achieve unless the vacuum energy is not subject to gravitation \citep{weinberg_de, ellis}. This theoretical difficulty motivated investigations of other possible explanations for the origin of the accelerated expansion. These are generically labelled {\em Dark Energy} although this term is a little reductive.

The simplest approach to explaining dark energy is certainly to go back to Einstein's initial cosmological constant but breaking the link with the quantum vacuum just by considering $\Lambda$ as a generic property of gravity. This is a very simple assumption that allows to explain the observational data although calling for a new constant is considered by many a kind of theoretical failure recognition. More general approaches involve modified gravity theories that could explain the observations in theories beyond general relativity.

The generic {\em Dark Energy} models (sometimes called {\em quintessence}) assume the presence of a scalar field, similar to the one that would be responsible for inflation. In such a case, the potential energy associated with this scalar is the origin of the dark energy density. The mass of the scalar field needs to be very small in order to prevent dark energy to cluster like matter. The fine tuning question of why the cosmological constant is sizable precisely when we can observe it can be solved in these models by a {\em tracker} behavior of dark energy: until matter-radiation equality, the dark energy density remains smaller than the matter density but follows its variation. It is only after matter-radiation equality that dark energy starts to behave as a cosmological constant. Another fine tuning issue of these models is that the mass of the scalar field needs to be high in order to have sizeable effects on cosmological scales, which is in contradiction with local tests of the equivalence principle. This can be overcome through the "Chameleon mechanism" \citep{khoury} in which the the mass of the field actually depends on the local density, so that locally the agreement with the the tests for the equivalence principle is preserved.
A generic signature of these models is that as the potential energy can vary in space--time, allowing for variations of the dark energy equation of state as a function of redshift. This motivates current experiments aiming at measuring $w_0$ and $w_a$, the two first orders of the expansion of the equation of state as a function of redshift. Observational data currently constrains $w_0=-1.08\pm0.15$ and $w_a=0.08\pm 0.81$ \citep{anderson} which neither contradicts nor supports dark energy models. Note that models with $w<-1$ are called "phantom energy" \citep{caldwell_phantom} and, in FLRW cosmology, lead to a infinite energy density for the dark energy within finite time, leading to a so-called "Cosmic Doomsday" where gravitational repulsion would end up overcoming all structures, including microscopic ones in a "Big Rip".

An appealing explanation to accelerated expansion is {\em back-reaction} \citep{buchert}. This relies on the fact that gravity is non-linear and that it is not obvious that solving the Einstein equations for an exactly homogeneous Universe and subsequently adding the inhomogeneities, as done within the FLRW paradigm, would give the same as solving the equations accounting from the beginning for the inhomogeneities. If the tiny primordial density perturbations have a sufficient effect to modify the average dynamics of the expansion, this could be an extremely elegant explanation to the observations without the need for modified gravity or actual dark energy. Unfortunately, calculating the magnitude of back-reaction is extremely difficult as it is precisely a non-linear effect for which perturbative approaches are not valid. There are no definitive conclusions regarding the viability of this explanation \citep{clarkson} as various authors find contradicting results using different approximations. A full treatment is unfortunately currently out of reach due to the extreme difficulty of such calculations.

\subsection{Dark Energy summary}
Similarly as with dark matter, one must recognize that although the evidence for accelerated expansion is strong, and despite the fact that the data is extremely well adjusted by a cosmological constant within the $\Lambda$CDM model, the profound nature of dark energy remains mysterious and makes the average cosmologist feel extremely uncomfortable. From the many independent probes pointing towards accelerated expansion, it seems unlikely that the final explanation will be an observational bug. The effect seems to be firmly established and therefore requires a theoretical explanation. There is no definitive reason for now to favor any of the approaches proposed until now: simple cosmological constant, modified gravity, dark energy or back-reaction.

An extremely important observational effort has been undertaken by the community to obtain more informations on dark energy, especially on the value and possible evolution of its equation of state. Such data could be decisive in the future years, especially if an excursion away from $(w_0=-1,w_a=0)$ is observed providing support to quintessence-like dark energy models.

\section{Known issues with the $\Lambda$CDM model}
Although very convincing from many points of views as discussed above, the $\Lambda$CDM model suffers from a few inconsistencies that should not be left aside. Of course the ignorance of the deep nature of the constituents of the model is certainly the major one: we still have no convincing or experimentally confirmed explanations for both Dark Matter and Dark Energy that are fitted with $\Lambda$CDM to represent 96\% of the energy content of the Universe. This is certainly a serious problem but could hopefully be overcome in the future by decisive observations. However, one could argue that this issue is somehow more an issue regarding the interpretation of the results of fitting data that is perfectly consistent with the $\Lambda$CDM model. 

However, there remain actual inconsistencies between observations and $\Lambda$CDM predictions. We have already discussed the Lithium-7 problem which is not found to be abundant enough in the Universe with respect to predictions from Big-Bang Nucleosynthesis theory. As said before, this result could be well explained by a bias in the measurements of the primordial Lithium-7 as it can be burnt in stars~\cite{iocco}. Another somewhat related issue is that of the lack of observation of population III stars, in the astrophysical jargon, this is the name of the first stars to have formed. We know that the stars around us burn their hydrogen through a complex  series of nuclear reactions that involve heavy nuclei such as C, N and O. These elements were not present in the early Universe as Big-Bang Nucleosynthesis did not produce anything heavier than Lithium. Such stars however need to have existed and their nuclear reactions need to have been efficient enough to produce massively the UV light that reionized the Universe and to have enriched the interstellar medium with heavy elements by their final supernova explosion. Current theory (although debated) favors the idea that such nuclear reactions only based on light nuclei are possible in extremely heavy ($\sim 200 M_\odot$) stars which would also explain why no population III was ever observed because stars as massive as this would have a short lifespan ($\sim 3$ million years) and end up in supernovae or black-holes (hence possibly forming quasars whose UV light contributed to reionization).
A possible illustration of the bias in measuring Lithium from stars and a possible link with population III stars comes from the recently observed star SDSS J102915+172927 \citep{star_nolithium} which exhibits a very low metal content\footnote{Although it is incorrect in the strict sense, in astrophysics all atoms  other than Hydrogen and Helium are labelled as metals.}, especially an amount of Lithium at least 50 times lower than produced during Big-Bang Nucleosynthesis. According to the standard nuclear evolution theory, the age of this star would be of order 13 billion years, making it the oldest star ever observed, and challenging the age of the Universe estimated within $\Lambda$CDM. However, this star has such a low mass (about $0.8 M_\odot$) that it should not have formed according to the stellar theory. The age estimation is therefore subject to large theoretical uncertainties and the low amount of Lithium shows that it can indeed be burnt in stars. Finally, most of the experts seem to agree that the existence of this star, the lack of population III stars and the Lithium issue do not seriously challenge $\Lambda$CDM but rather illustrate complex processes still to be understood in details.

The most severe contradictions between $\Lambda$CDM predictions and observations arise in the small scale sector, more precisely in the regime where gravity is strongly non-linear and requires using complex N-body and hydrodynamical numerical simulations. For instance, in the "bullet cluster", the velocity of the shock between the two colliding clusters is measured from the X-ray observation of the gas shock to be $\sim 3000~\mathrm{km.sec^{-1}}$ \citep{mastropietro}, a value that is well over expectations from numerical simulations in $\Lambda$CDM that rarely exceed $\sim 1800~\mathrm{km.sec^{-1}}$ \citep{leekomatsu}. At even smaller scales, when describing galactic dynamics, several disagreements have been pointed out between $\Lambda$CDM numerical simulations and observations \cite{mondreview}:
\begin{itemize}
\item The number of satellite galaxies is observed to be an order of magnitude smaller in our local group than in the simulations \citep{kravtsov}. This could be explained if the simulations do not properly account for the tidal mass loss in the sub-halo structure during the star formation, preventing small sub-halos from becoming actual galaxies. It is not clear for now if this discrepancy reveals a real $\Lambda$CDM problem or if it only shows that numerical simulations are not accurate enough at these extreme scales.
\item We already mentioned the Tully-Fisher relation that does not find a natural explanation in $\Lambda$CDM while it is obtained in a straightforward manner in MOND.
\item The most central regions of the galaxies are observed to have a rather flat density profile while numerical simulations predict a strong cusp at the center. Again, this could be explained by a lack of resolution and missing processes in the simulations, such as bulk gas motions induced by stellar feedback (supernovae explosions) that could flatten the central profile of the galaxies \citep{mashchenko}.
\end{itemize}

Several claims of anomalies in the Cosmic Microwave Background have been made in the recent years from very attentive observations of the WMAP data (although the WMAP team itself denies the statistical significance of these anomalies \citep{wmap_anomalies}). The low amplitude of the CMB quadrupole with respect to theoretical expectations has been subject to many publications (\citep{copi,doc} for instance). The statistical significance of this low quadrupole is however very small if one considers the large cosmic variance associated with this measurement (only five modes at $\ell=2$ are available to measure the variance of this multipole) and cannot be considered as a detection of  a departure from $\Lambda$CDM as it is perfectly consistent with a statistical fluctuation. As in any set of noisy measurements, one expects around a third of them to be at more than one standard deviation from the expected value, and about 5\% at more than two standard deviations. The correct statistical significance of other anomalies in the CMB (such as the so called "Axis of evil" of aligned multipoles \citep{doc, schwarz, copi2} or hemisphere asymmetry \citep{eriksen, freeman, hansen}) is probably even smaller and in any case very difficult to assess rigorously as they correspond to {\em a posteriori} cuts performed on the data. As clearly analyzed in \cite{bunn_stat} one cannot take at face value the statistical significance obtained by observing some unusual feature in a large dataset and then {\em a posteriori} deriving the statistical significance of this feature. Penalty factors reducing (usually a lot) the {\em a posteriori} significance should be included in order to account for the number of cuts on the data that were needed to identify this feature, the various choices of combinations of the data that were looked at and so on\footnote{The problem of the choice of the cuts made on a dataset in order to claim for a discovery is a serious one. If the cuts are decided from the data itself, there is an important risk to perform "educated cuts" that tend to maximize the significance of the effects that are searched for. This is the reason why some teams prefer to perform "blind analysis" where all the cuts are decided on simulated data before looking at the data, and then applied strictly on the data without modification.}. These penalty factors are impossible to estimate accurately but would clearly reduce to almost nothing the statistical significance of the mentioned features. The example of "S" and "H" letters found on WMAP CMB maps is a good one with this respect: it is not surprising to be able to find structures resembling letters in a map consisting of random fluctuations if the letters are not predicted prior to observations. The actual significance of these initials is completely different between the (real) case where they were not predicted before looking at the map ({\em a posteriori} observation) or the (unreal) case where such an observation would have been predicted before (hence {\em a priori}). All of these so-called "anomalies" in the CMB must therefore be considered with extreme care as they illustrate the difficulty of assessing statistical significance on details observed within large data set. Our human brain is trained to pick up such details but their actual importance in terms of statistics needs to be rigorously estimated in order to provide actual information on the validity on the model. Trying to find such anomalies is however a very interesting perspective as it may eventually reveal some significant inconsistencies although it has not been the case up to now.

It appears that most of the remaining disagreements between $\Lambda$CDM predictions and observations could be explained by a lack of details in the modeling of complex processes at the smallest scales. These are hard to control in numerical simulations. Hopefully, the increasing computing power will soon allow one to achieve more accurate numerical simulations and test if these discrepancies still remain. It is important to remark that no disagreement is left in the broad, large scale description of the Universe, while the independent observations have flourished in the last few years, giving increased confidence in the validity of the $\Lambda$CDM paradigm.

\section{Conclusions}
We have reviewed observational results supporting the idea that the observable region of our Universe is well described by the so-called $\Lambda$CDM model, a particular model within Friedman-Lemaître-Robertson-Walker cosmologies where the geometry is flat and as much as 70\% of the energy content is in the form of well measured but still unexplained dark energy, and 25\% in the form of dark matter, also well measured but lacking direct explanation.

The most important assumption at the basis of the FLRW cosmologies, the Cosmological Principle, is more and more verified with increasing dataset allowing to map the Universe on large scales. The isotropy is well established thanks to the exquisite temperature uniformity of the Cosmic Microwave background and large galaxy redshift surveys do not exhibit structure beyond the clusters and filaments that are expected (from numerical simulations) in $\Lambda$CDM at those scales. The Copernican principle is not as well established, but increasingly accurate measurements, such as the lack of significant kinetic Sunyaev-Zel'dovitch effect allow to rule out inhomogeneity up to most of the size of the observable Universe. The application of the Cosmological Principle therefore seems well justified by the observations.

The idea of a Universe that started in a very dense and hot Big-Bang is also supported by long running observations: the redshift of the galaxies supports the idea of expansion, the existence of the Cosmic Microwave Background and the agreement between light elements abundances and Big-Bang Nucleosynthesis are both evidence for a hot and dense past where the cooling due to expansion triggered phase transitions leaving observable relics. Reionization, although not as well explored up to now, also favors the same scenario.

The observation of the tiny fluctuations in the Cosmic Microwave Background temperature and polarization allowed cosmology to enter in an era of precision measurements, all of them showing an excellent matching with theoretical predictions: baryonic acoustic oscillations in the CMB angular power spectrum, perfect agreement between cosmological parameters measured with the CMB and other probes (such as with Big-Bang Nucleosynthesis for the baryonic matter content). Supporting arguments for an inflationary early Universe were also brought by the CMB observations: the matching between peaks and troughs in the E and T power spectra favors adiabatic primordial fluctuation, the spectral index for primordial fluctuations agrees well with inflation predictions and the CMB fluctuations are in excellent agreement with the predicted Gaussian statistics.

The dominance of dark non-baryonic matter with respect to ordinary luminous matter is also strongly supported by many observations. On galactic scales the stellar rotation curves are well explained by dark matter while on the larger scales the cluster dynamics require massive amounts of dark matter. This dark matter is even mapped in the clusters through the observation of weak lensing showing large clumps at the center of the clusters that seem to interact weakly when clusters collide. This is in itself the description of dark matter: massive but weakly interacting. Similarly on even larger scales, the observation of the baryonic acoustic oscillations in the galaxy distribution can only be explained by dark matter. On the largest scales, numerical simulations need dark matter to be able to reproduce the large scale structure we observe. All of these observational results are well explained by weakly interacting massive particles, possibly relic of supersymetry, that unfortunately still lacks direct detection be it through direct production in the colliders or direct detection in underground laboratories. Of course, other possible explanations are debated in the community, such as modified gravity. For now, although such models are successful in explaining the small scales observations attributed to dark matter, they do not succeed in explaining cluster dynamics, baryonic acoustic oscillations or the low baryon content (from CMB or Big-Bang Nucleosynthesis) with respect to the 30\% overall matter content suggested by many different probes. The dark matter question is however still largely open and more observations are required to shed light on this mystery.

The apparent main component of our Universe is also still mysterious. Type Ia supernovae brought strong evidence for an accelerated expansion of the Universe than can be explained within FLRW cosmologies by 70\% of the energy content of the Universe in the form of a negative pressure cosmological constant. Various probes including baryonic acoustic oscillations and Cosmic Microwave Background converge towards the same flat, cosmological constant dominated Universe. From the observational point of view, the situation is therefore rather clear: the history of expansion is consistent with the presence of the cosmological constant. The theoretical explanation is however far from being clear: the identification of the cosmological constant with the particle physics vacuum energy is problematic and more general dark energy models, although compatible with the data cannot be distinguished for now from the cosmological constant. The latter could be incorporated as a simple gravitation property (a new constant) but this explanation doesn't satisfy most of the theorists. Back-reaction, the gravitational effect of density fluctuations on the expansion history could provide an elegant explanation to the apparent acceleration of the Universe if including them modifies significantly the FLRW model. Unfortunately, these effects are difficult to calculate and no conclusion on the importance of back-reaction can be drawn as of today. The domination of the Universe by an unexplained fluid with strange properties (negative pressure) is undoubtedly one of the major problems in cosmology today and intense experimental and theoretical efforts are undertaken to solve it.

We have also discussed several contradictions between predictions of $\Lambda$CDM and observations and have shown that they mainly concern quantities which are difficult to measure in an unbiased manner (Lithium-7) or hard to calculate theoretically. This is especially true for the small (galactic scale) dynamics whose predictions heavily rely on numerical simulations at the limit of their resolution while in the broad picture, no significant discrepancy remain.

There are also important unresolved theoretical issues in $\Lambda$CDM. One of the most important ones is certainly related to inflation. While inflation is usually considered as the best model to produce primordial fluctuations with the desired scale-invariant spectrum, the actual mechanism producing the transition from quantum to classical fluctuations is not understood. Other theoretical issues are strongly related to deep philosophical questions. The haunting one is obviously the question of the start of the Universe. We know that the idea of a Big Bang is rather naive in the sense that it is an extrapolation down to $t=0$ of a model based on a theory that breaks down at least at the Planck time (because at such densities ones needs an eagerly awaited quantum theory of gravitation). The $\Lambda$CDM model, only accurate after the Planck time, can therefore thankfully reject the question to its outside. The question of the start of the Universe (if any) nevertheless remains among the most important ones and is not addressed in a satisfying manner by cosmology. Another open theoretical question, heavily related to a profound philosophical issue, is that of what lies beyond our horizon. Applying in a strict manner the $\Lambda$CDM model in the case of flat or open Universe (hence spatially infinite), or following the predictions of chaotic inflation, lead to the multiverse hypothesis \citep{tegmark_multiverse} that seems hard to test and, although seducing from some points of views cannot be considered without major philosophical consequences. 

It would certainly be way to daring to consider $\Lambda$CDM as the ultimate model of cosmology, dark matter and dark energy are so compelling issues that a victorious attitude would be out of purpose. On the other hand, the systematic agreement between new and increasingly accurate observations and $\Lambda$CDM predictions is very impressive and explains the fact that most of the community considers $\Lambda$CDM as the standard model for cosmology. It is certainly the best model we have to explain the observations although it leaves a number of deep questions widely opened, mostly concerning the deep nature of the constituents of the Universe.

\section*{Acknowledgements}
The author wishes to thank the organizers of the "Philosophical Aspects of Modern Cosmology" workshop for such a friendly, pleasant and illuminating workshop. He also wishes to thank Henrik Zinkernagel, Nicolás Busca, Jean Kaplan and Cécile Roucelle for fruitful discussions and constructive comments on the manuscript. The author also thanks very warmly the two anonymous referees who provided constructive comments and suggestions that improved this article.


\begin{thebibliography}{00}
\bibitem[Adriani et al.; 2008]{pamela} O. Adriani et al., Nature 458, 607 (2009).
\bibitem[Ahmed et al.; 2010]{cdms} Z. Ahmed et al., Science 327, 1619, (2010).
\bibitem[Alcock et al.; 1993]{alcock} Ch. Alcock et al., Nature 365, 623 (1993).
\bibitem[Alpher et al; 1948]{alpher} R.A. Alpher, H.A. Bethe and G. Gamow, Phys. Rev. 73, 803 (1948).
\bibitem[Alpher and Herman; 1948]{alpher_herman} R.A. Alpher and R.C. Herman, Phys. Rev. 74 (12): 1737–1742 (1948).
\bibitem[Anderson et al.; 2012]{anderson} L. Anderson et al., MNRAS 427 (4): 3435-3467 (2012), {\tt arXiv:1203.6594}.
\bibitem[Aprile et al.; 2011]{xenon} E. Aprile et al., Phys. Rev. Lett. 107, 131302 (2011) {\tt  arXiv:1104.2549}.
\bibitem[Armengaud; 2011]{edelweiss} É. Armengaud et al., Physics Letters B 702, 329-335 (2011), {\tt arXiv:1103.4070}.
\bibitem[Aubourg et al.; 1999]{aubourg_eros} É. Aubourg et al., A\&A 347, 850 (1999).
\bibitem[Aubourg; 2011]{aubourgHDR} É. Aubourg, Habilitation à diriger des Recherches, Université Paris-Diderot, 2011.
\bibitem[Barger et al.; 2009]{barger} V. Barger et al., Phys.Lett.B678:283-292 (2009).
\bibitem[Baumann et al.; 2008]{baumann} D. Bauman et al., {\tt 0811.3919v2}
\bibitem[Becker et al.; 2001]{becker} Astronomical Journal 122 (6): 2850–2857 (2001), {\tt arXiv:astro-ph/0108097}.
\bibitem[Bernabei et al.; 2008]{dama} R. Bernabei et al., European Physical Journal C, 56, 333 (2008).
\bibitem[Beutler et al.; 2011]{beutler} F. Beutler, et al., MNRAS, 416, 3017 (2011)
\bibitem[Bennet et al.; 2010]{wmap_anomalies} C. L. Bennett et al.,  (2010) {\tt arXiv:1001.4758}.
\bibitem[Beno\^{i}t et al.; 2003]{archeops} Beno\^{i}t A. et al. 2003, A\&A, 399, No. 3, L19.
\bibitem[Blake et al.; 2011]{blake} C. Blake et al., MNRAS, 415, 2892 (2011)
\bibitem[Bond and Efstathiou; 1987]{bond_efstathiou} J.R. Bond and G. Efstathiou, MNRAS 226, 655-687 (1987).
\bibitem[Bondi; 1947]{ltb} H. Bondi, Mon. Not. Roy. Astron. Soc. 107, 410 (1947).
\bibitem[BOSS Press release; 2011]{BOSS} BOSS Press release {\tt http://www.sdss3.org/press/20110111.largestimage.php}
\bibitem[Bovy and Tremaine; 2012]{bovy} J. Bovy and S. Tremaine, ApJ 756,1 (2012) , {\tt arXiv:1205.4033}.
\bibitem[Buchert; 1999]{buchert}T. Buchert, Gen. Rel. Grav. 32 (2000) {\tt arXiv:gr-qc/9906015}.
\bibitem[Bunn and Hogg; 2009]{bunn} E.F. Bunn and D.W. Hogg, Am.J.Phys.77:688-694,2009, {\tt arXiv:0808.1081}.
\bibitem[Bunn; 2010]{bunn_stat} E.F. Bunn, Proceedings of the 2010 "rencontres de Moriond", {\tt arXiv:1006.2084}
\bibitem[Busca et al,;2012]{busca} N.G. Busca et al., submitted to A\&A, DOI: 10.1051/0004-6361/201220724, {\tt arXiv:1211.2616}.
\bibitem[Caffau et al.; 2011]{star_nolithium} E. Caffau et al., Nature, Volume 477, Issue 7362, pp. 67-69 (2011).
\bibitem[Calchi Novati et al.; 2005]{agape} S. Calchi Novati et al., A\&A, vol. 443, no. 3, pp. 911–928 (2005).
\bibitem[Caldwell et al.; 2003]{caldwell_phantom} R.R. Caldwell, M. Kamionkowski, N.N. Weinberg, Phys.Rev.Lett. 91, 071301 (2003), {\tt arXiv:astro-ph/0302506v1}
\bibitem[Caldwell and Stebbins; 2008]{caldwell_stebbins} R. R. Caldwell and A. Stebbins, Phys. Rev. Lett. 100 (2008) 191302 {\tt arXiv:0711.3459}.
\bibitem[Chaboyer et al.; 1996]{chaboyer} B. Chaboyer et al., Science 271, 957 (1996).
\bibitem[Clarkson et al.; 2011]{clarkson} C. Clarkson, G. Ellis, J. Larena, O. Umeh, Rept.Prog.Phys. 74, 112901 (2011), {\tt arXiv:1109.2314}.
\bibitem[Clowe; 2006]{clowe} D. Clowe et al., ApJ 648, L109 (2006).
\bibitem[Cole et al.; 2005]{cole2dF} S. Cole et al., MNRAS, 362, 505 (2005).
\bibitem[Colin et al; 2011]{colin} J. Colin et al, Mon. Not. Roy. Astron. Soc. 414 (2011) 264-271 {\tt arXiv:1011.6292}.
\bibitem[Copi et al.; 2004]{copi2} C. J. Copi, D. Huterer and G. D. Starkman, Phys. Rev. D 70, 043515 (2004).
\bibitem[Copi et al.; 2007]{copi} C. J. Copi et al., Phys. Rev. D 75, 023507 (2007).
\bibitem[de Jong et al.; 2006]{crotts} J.T.A. de Jong et al., A\&A, vol. 446, no. 3, pp. 855–875 (2006).
\bibitem[de Oliveira-Costa et al.; 2004]{doc} A. de Oliveira-Costa et al., Phys. Rev. D 69, 063516 (2004).
\bibitem[Dicke et al.; 1965]{dicke} Dicke, R. H., Peebles, P. J. E., Roll, P. G., Wilkinson, D. T., Astrophysical Journal, vol. 142, p.414-419 (1965).
\bibitem[Efstathiou et al.; 1990]{efstathiou} G. Efstathiou, W. Sutherland and S.J. Maddox, Nature 348, 705 (1990).
\bibitem[Eisenstein et al.; 2005]{eisenstein} D. Eisenstein et al., ApJ, 633, 560 (2005).
\bibitem[Ellis et al.; 2010]{ellis} G.F.R. Ellis, H. van Helst, J. Murugan, J.-Ph. Uzan, Class.Quant.Grav. 28, 225007 (2011), {\tt arXiv:1008.1196}.
\bibitem[Eriksen et al.; 2004]{eriksen} H. K. Eriksen et al., Astrophys. J. 605, 14 (2004).
\bibitem[Famaey and McGaugh; 2011]{mondreview} B. Famaey and S. McGaugh, Living Reviews in Relativity 15 (2012).
\bibitem[Freedman et al; 2001]{freedman} W. Freedman et al, Astrophys.J.553:47-72, 2001, {\tt  astro-ph/0012376}.
\bibitem[Freeman et al.; 2006]{freeman} P. E. Freeman et al., Astrophys. J. 638, 1 (2006).
\bibitem[Gabrielli et al.; 2005]{gabrielli} A. Gabrielli et al., Statistical Physics for Cosmic Structures, Springer Verlag, Berlin (2005).
\bibitem[Gamow; 1948a]{gamow} Phys. Rev. 74, 505–506 (1948).
\bibitem[Gamow; 1948b]{gamow1} G. Gamow,  Nature 162 (4122): 680–682 (1948).
\bibitem[Granett et al.; 2009]{granett} B.R. Granett et al., Astrophys.J. 701 (2009), {\tt arXiv:0812.1025}
\bibitem[Gunn and Peterson; 1965]{gunnpeterson} J.E. Gunn and B.A. Peterson, ApJ 142: 1633–1641 (1965).
\bibitem[Hanany et al.; 2002]{maxima} Hanany, S. et al. 2000, ApJ, 545, L5.
\bibitem[Hansen et al.; 2009]{hansen} F. K. Hansen et al., Astrophys. J. 704, 1448 (2009).
\bibitem[Hogg et al.; 2005]{hogg} D.W. Hogg et al., ApJ, 624, 54 (2005) {\tt arXiv:astro-ph/0411197}
\bibitem[Hubble; 1929]{hubble} Hubble, E. P. (1929) Proc. Natl. Acad. Sci. USA 15, 168–173
\bibitem[Iocco et al.; 2009]{iocco} F. Iocco et al., Phys.Rept.472:1-76,2009, {\tt arXiv:0809.0631}.
\bibitem[Jeans; 1922]{jeans} J.H. Jeans, MNRAS 82 (1922).
\bibitem[Jee et al.; 2012]{jee} M.J. Jee et al., ApJ, 747, 96 (2012), {\tt arXiv:1202.6368}.
\bibitem[Jones et al.; 2004]{jones} B.J.T. Jones et al., Rev. Mod. Phys. 76,  1211 (2004) {\tt arXiv:astro-ph/0406086v1}.
\bibitem[Kapteyn; 1922]{kapteyn} J. Kapteyn, Astrophysical Journal 55 (1922).
\bibitem[Kellogg et al.; 1972]{kellogg} E.M. Kellogg et al., Astrophys. Journ. Lett. 174, L65 (1972).
\bibitem[Khoury and Weltman; 2003]{khoury} J. Khoury and A. Weltman, Phys. Rev. D 69, 044026 (2004), {\tt arXiv:astro-ph/0309411}
\bibitem[Kirkman et al.; 2003]{bbn_ob} D. Kirkman et al., Astrophys.J.Suppl. 149 1 (2003), {\tt  arXiv:astro-ph/0302006}.
\bibitem[Komatsu et al.; 2011]{wmap7_komatsu} Komatsu, E., et.al., 2011, ApJS, 192, 18, {\tt arXiv:1001.4538}.
\bibitem[Kovac et al.; 2002]{dasi} J. Kovac et al., Nature 420 (2002) 772.
\bibitem[Kowalski et al.; 2008]{kowalski} M. Kowalski et al., Astrophys. J. 686, 749 (2008), {\tt arXiv:0804.4142v1}.
\bibitem[Kravtsov et al.; 2004]{kravtsov} A. Kravtsov et al., Astrophys.J 609 (2004).
\bibitem[Landau et al.; 2012]{landau} S. Landau, C. Scóccola, D. Sudarsky, Phys. Rev. D 85, 123001 (2012), {\tt arXiv:1112.1830}.
\bibitem[Lee and Komatsu; 2010]{leekomatsu} J. Lee and E. Komatsu, Astrophys. J, 718 (2010), {\tt arXiv:1003.0939}.
\bibitem[Luzzi et al; 2009]{luzzi} Luzzi, G., Shimon, M., Lamagna, L., et al. 2009, ApJ, 705, 1122.
\bibitem[Maartens; 2011]{maartens} R. Maartens, Phil. Trans. R. Soc. A 369 (2011) 5115-5137, {\tt  arXiv:1104.1300}.
\bibitem[Mahdavi et al.; 2007]{mahdavi} A. Mahdaviet al., ApJ, 668, 806 (2007), {\tt arXiv:0706.3048}.
\bibitem[Martinez and Trimble; 2008]{martineztrimble} V.J. Martinez and V. Trimble, proc. of {\em Cosmology across Cultures}, Granada, 2008, {\tt arXiv:0904.1126}.
\bibitem[Mashchenko et al.; 2006]{mashchenko} S. Mashchenko et al., Nature 442 (2006).
\bibitem[Mastropietro and Burkert; 2008]{mastropietro} C. Mastropietro and A. Burkert, MNRAS, 389 (2008).
\bibitem[Mather et al.; 1996]{mather} J. Mather et al., Astrophysical Journal v.473, p.576 (1996).
\bibitem[Milgrom; 1983]{mond} M. Milgrom et al., ApJ 270, 365 (1983).
\bibitem[Moni Bidin et al.; 2012]{bidin}C. Moni Bidin et al., ESO Press Release \#1217 (2012), {\tt arXiv:1204.3924}
\bibitem[Moffat; 2011]{moffat} J.W. Moffat, Conference {\em Two Cosmological models}, Mexico (2010), {\tt arXiv:1101.1935}.
\bibitem[Nadathur and Sarkar; 2010]{sarkar2010} S. Nadathur and S. Sarkar, Phys.Rev.D83:063506 (2011) {\tt arXiv:1012.3460}.
\bibitem[Netterfield et al.; 2002]{boomerang} Netterfield, C. B. et al. 2002, ApJ, 568, 38.
\bibitem[Noterdaeme et al.; 2011]{noterdaeme} P. Noterdaeme et al., A\&A 526, L7 (2011) {\tt arXiv:1012.3164}.
\bibitem[Olive; 2012]{olive} K. Olive, Proc. 7th DSU Conference, Beijing, China (2012), {\tt  arXiv:1202.2324}.
\bibitem[Oort; 1932]{oort} J. Oort, , Bull. Astron. Inst. Netherlands, 6, 249 (1932).
\bibitem[\"Opik; 1915]{opik} E. \"Opik, Russian Astronomical Society Bulletin, 21, 150 (1915).
\bibitem[Penzias and Wilson; 1965]{penzias_wilson} A. Penzias and R. Wilson, Astrophysical Journal, vol. 142, p.419-421 (1965).
\bibitem[Perez et al.; 2006]{perez} A. Perez, H. Sahlmann and D. Sudarsky, Class.Quant.Grav. 23, 2317-2354  (2006), {\tt arXiv:gr-qc/0508100}
\bibitem[Perlmutter et al.; 1999]{perlmutter} S. Perlmutter et al., Astrophys. J. 517, 565 (1999).
\bibitem[Phillips; 1993]{phillips} M.M. Phillips, Astrophys. J. 413, 105 (1993).
\bibitem[Pietronero et al.; 2000]{pietronero} L. Pietronero and F. Sylos Labini, proc. 7th Course in astrofundamental physics, Erice (1999), {\tt arXiv:astro-ph/0002124}.
\bibitem[Readhead et al.; 2002]{cbi} A.C.S. Readhead et al., Science 306 (2002) 836.
\bibitem[Reid; 2012]{reid} B. Reid et al., MNRAS 426, 4 (2012), {\tt arXiv:1203.6641}.
\bibitem[Riess et al.; 1998]{riess} A.G. Riess et al., Astron. J. 116, 1009 (1998).
\bibitem[Rubin et al.; 1980]{rubin} V. Rubin et al, Astronophysical Journal, 238:471 (1980).
\bibitem[Rugh and Zinkernagel; 2002]{rugh_zinkernagel} S.E. Rugh and H. Zinkernagel, SHPMP vol. 33, 663 (2002), {\tt arXiv:hep-th/0012253}.
\bibitem[Sanchez et al.; 2012]{sanchez} A. Sanchez et al., MNRAS 425, 1 (2012), {\tt arXiv:1203.6616}.
\bibitem[Schwarz et al.; 2004]{schwarz} D. J. Schwarz et al., Phys. Rev. Lett. 93, 221301 (2004).
\bibitem[Scrimgeour et al.; 2012]{scrimgeour} M. Scrimgeour et al., MNRAS 425, 116-134 (2012) {\tt arXiv:1205.6812}.
\bibitem[Sherwin et al.; 2011]{sherwin} B.D. Sherwin et al., PRL 107, 021302 (2011), {\tt arXiv:1105.0419}.
\bibitem[Smoot et al.; 1992]{smoot} G.F. Smoot et al., Astrophysical Journal v.396, p.1 (1992).
\bibitem[Stebbins; 2007]{stebbins} A. Stebbins, {\tt arXiv:astro-ph/0703541}.
\bibitem[Sunyaev and Chubla; 2009]{sunyaev} R.A. Sunyaev and J. Chubla, Astron. Nachr. / AN 330, No. 7, 657-674 (2009).
\bibitem[Sunyaev and Zel'dovitch; 1972]{sz1} R. A. Sunyaev, and Y. B. Zeldovich, Comments on Astrophysics and Space Physics, 4, 173 (1972)
\bibitem[Sunyaev and Zel'dovitch; 1980]{sz2} R. A. Sunyaev and Y. B. Zeldovich, MNRAS, 190, 413 (1980).
\bibitem[Tegmark; 2009]{tegmark_multiverse} M. Tegmark, In "Universe or Multiverse?", B. Carr ed. Cambridge University Press (2007), {\tt arXiv:0905.1283}.
\bibitem[Tisserand et al.; 2007]{eros} P. Tisserand et al., A\&A 469, 387 (2007).
\bibitem[Tully and Fisher; 1977]{tullyfisher} R.B. Tully and J.R. Fisher, A\&A 54, 661 (1977).
\bibitem[Uzan et al.; 2008]{uzan}  J.-P. Uzan, C. Clarkson and G.F.R. Ellis, Phys. Rev. Lett. 100 (2008) 191303 {\tt arXiv:0801.0068}.
\bibitem[Weinberg et al.; 2012]{weinberg}D.H. Weinberg et al., Physics Reports (2012), {\tt arXiv:1201.2434}.
\bibitem[Weinberg; 1989]{weinberg_de}S. Weinberg, Reviews of Modern Physics, Volume 61, Issue 1 (1989).
\bibitem[Zhang and Stebbins; 2010]{zhang_stebbins} P. Zhang and A. Stebbins, Phys. Rev. Lett. 107 (2011) 041301 {\tt arXiv:1009.3967}.
\bibitem[Zwicky; 1933]{zwicky} F. Zwicky, Helv. Phys. Acta, 6, 110 (1933).
\bibitem[Zwicky; 1937]{zwicky2} F. Zwicky, ApJ 86, 217 (1937).
\end{thebibliography}
\end{document}